\documentclass[%
 amsmath,amssymb,
 superscriptaddress,
 twocolumn,
 pra
]{revtex4-2}
\usepackage{graphicx,amsfonts,amssymb,amsmath,hyperref}
\usepackage{dcolumn}
\usepackage{textcomp}
\usepackage{mathtools}
\usepackage{esint}
\usepackage{braket}
\usepackage{lineno}
\usepackage{bm}
\usepackage[normalem]{ulem}
\usepackage[dvipsnames]{xcolor}

\begin{document}
\newcommand{\chcom}[1]{{\color{blue}CH: #1}}
\newcommand{\aacom}[1]{{\color{green} #1}}

\title{Bidimensional measurements of photon statistics within a multimodal temporal framework}

\author{C.~Hainaut}
\email{clement.hainaut@univ-lille.fr}
 \affiliation{Univ. Lille, CNRS, UMR 8523–PhLAM–Physique des Lasers, Atomes et Molécules, Lille, France\\}
 
\author{K.~Ouahrouche}%
\affiliation{Univ. Lille, CNRS, UMR 8523–PhLAM–Physique des Lasers, Atomes et Molécules, Lille, France\\}

\author{A.~Ran\c{c}on}%
\affiliation{Univ. Lille, CNRS, UMR 8523–PhLAM–Physique des Lasers, Atomes et Molécules, Lille, France\\}
\affiliation{Institut Universitaire de France}

\author{G.~Patera}%
\affiliation{Univ. Lille, CNRS, UMR 8523–PhLAM–Physique des Lasers, Atomes et Molécules, Lille, France\\}
 
\author{C.~Ouarkoub}%
\affiliation{Univ. Lille, CNRS, UMR 8523–PhLAM–Physique des Lasers, Atomes et Molécules, Lille, France\\}

\author{M.~Le~Parquier}%
\affiliation{Univ. Lille, CNRS, UMR 8523–PhLAM–Physique des Lasers, Atomes et Molécules, Lille, France\\}

\author{ P.~Suret}%
\affiliation{Univ. Lille, CNRS, UMR 8523–PhLAM–Physique des Lasers, Atomes et Molécules, Lille, France\\}

\author{A.~Amo}%
\affiliation{Univ. Lille, CNRS, UMR 8523–PhLAM–Physique des Lasers, Atomes et Molécules, Lille, France\\}

\date{\today}

\begin{abstract}
Ultrafast imaging of photon statistics in two dimensions is a powerful tool for probing non-equilibrium and transient optical phenomena, yet it remains experimentally challenging due to the simultaneous need for high temporal resolution and statistical fidelity. In this work, we demonstrate spatially resolved single-shot measurements of photon number distributions using difference-frequency generation (DFG) in a nonlinear BBO crystal. We show that our platform can discriminate between coherent and thermal photon statistics across two spatial dimensions with picosecond resolution. At the same time, we find that the retrieved distributions deviate from the ideal ones, a consequence of vacuum contamination and the multimodal response of the amplifier. To explain this, we develop a temporal mode decomposition framework that captures the essential physics of signal amplification and fluorescence, and quantitatively reproduces the experimental findings. This establishes a robust approach for measuring two-dimensional photon statistics while clarifying the fundamental factors that limit the fidelity of such measurements.
\end{abstract}
\maketitle

\section{Introduction}
For decades, researchers have sought the ability to faithfully image spatial bi-dimensional information generated at incredibly fast timescales. Ultrafast chaotic behaviors, complex nonlinear optical phenomena, and rapid biological processes
like protein rearrangement are just a few examples of events that occur on timescales ranging from picoseconds to femtoseconds. Additionally, transient events, specific to two-dimensional systems, reflect many major fundamental mechanisms occurring in biology, physics, or chemistry \cite{Bard_2001_book,Gorkhover2016,Imada1998}.

Traditionally, pump-probe methods \cite{Hockett2011,Wong2012,Acremann2000,Feurer2003,Kodama2001,Velten2012,Barty2008,Hajdu2000,Zewail1988} have been used to capture these fast dynamics through repeated measurements. However, many ultrafast phenomena are either non-repetitive or challenging to reproduce \cite{Jalali2010,Solli2007,Poulin2006,Tuchin2007,iaulys2014,Sciamanna2015}. Significant efforts have been made to create platforms capable of performing ultrafast single-shot optical measurements \cite{Liang2018,Gao2014,Jin2024}, intending to be free of the paradigmatic repeatable measurement schemes. The main objective of these platforms is the possibility of measuring bi-dimensional photonic fields, at an ultrafast scale (picosecond to femtosecond information with high repetition rate) without alteration of the information of interest. Well-established strategies consist of using cutting-edge pulsed laser technologies to probe a system with strong optical pulses to have enough signal-to-noise ratio in a single-shot experiment and then retrieving the information of interest with various elaborated methods \cite{Wang2014,Abramson1978,Li2014,Nakagawa2014,Yue2017,Ehn2017}

An important strategy to realize ultrafast sampling and perform single-shot measurements is based on the use of highly nonlinear optical processes such as Sum-Frequency Generation (SFG) or Difference-Frequency Generation (DFG), driven by a strong ultrashort pulse in bulk nonlinear materials. Both processes convert the information contained in a 2D optical field to another wavelength and sample it on an ultrafast timescale determined by the pulse duration. The key distinction between the two methods lies in their respective properties: DFG can amplify the signal but introduces a fluorescence background, while SFG is background-free without vacuum contamination when reaching 100$\%$ conversion efficiency \cite{Eckstein2011}. As a result, SFG is typically preferred for relatively strong input signals, where the signal significantly exceeds the dark count and readout noise of conventional detectors. In contrast, DFG is better suited for weak input signals due to its gain capabilities. While DFG-based bi-dimensional imaging has been demonstrated  \cite{Devaux1995,Vaughan11}, no focus has been set on the influence of the nonlinear effects and vacuum contamination on the spatial statistical distributions of photon counts in single-shot images. A system capable of extracting this information with high fidelity would enable the evaluation of bi-dimensional intensity correlation functions, providing additional information on ultrafast 2D phenomena.

In this study, we explore these questions by using parametric down-conversion to perform time sampling within a picosecond window of a two-dimensional image. Through a quantitative analysis of the system response, we demonstrate that the concept of temporal modes, originating from quantum optics and describing the response of nonlinear pulsed systems, serves as an effective framework to understand the system's behavior. We further analyze the behavior of amplified vacuum fluctuations (i.e., fluorescence) and examine how known input 2D signals with distinct photon statistical distributions are altered by the DFG process. Finally, we provide detailed experimental analysis, theory and numerical investigation to explain quantitatively why the measurements deviate from the statistics of the input field due to fundamental vacuum contamination as well as due to the multimodal temporal response of the system. We obtain predictions that match the experimental data remarkably. This work provides a comprehensive analysis of the 2D photon distribution imaging framework using the DFG scheme. 

The paper is organized as follows. In the first part, we present two-dimensional single-shot measurements of photon statistics using coherent and thermal input states, showing that the system is capable of faithfully imaging complex spatial patterns while preserving qualitative differences between the two types of photon states. We then highlight the quantitative deviations of the measured statistics from the ideal photon distributions and identify their origin in vacuum contamination and the multimodal response of the nonlinear amplifier. In the second part, we introduce a theoretical and numerical framework based on temporal mode decomposition, which provides a quantitative understanding of the system's response. Starting from a simplified plane-wave configuration, we validate the multimodal description by comparing its predictions with experiments performed with vacuum, coherent, and thermal input states. Finally, we extend this approach to the two-dimensional configuration, demonstrating that the model accurately accounts for the measured distributions, which establishes a comprehensive framework for DFG-based ultrafast-photon statistics imaging.

\section{2D photon statistics measurements}

The system under study consists of a bulk material with second-order nonlinearity $\chi^{(2)}$, namely, a Beta Barium Borate (BBO) crystal of length $l_c=2$~mm, interacting with two optical fields. A strong picosecond pulsed field ($\approx$ 0.1 mJ per pulse), referred to as the pump, with a wavelength $\lambda_{\text{p}}=400$ nm ($\omega_{\text{p}}=2\pi\times 749.2$ THz) and a continuous-wave signal at $\lambda_{\text{s}}=840.1$ nm ($\omega_{\text{s}}=2\pi\times 356.7$ THz) are spatially overlapped within the crystal. By appropriately selecting the phase-matching angle of the crystal, part of the pump field at $400$~nm converts and amplifies the signal field at $\lambda_{\text{s}}=840.1$ nm and an idler field centered at $\lambda_{\text{i}}=763.3$ nm ($\omega_{\text{i}}=2\pi\times 392.5$ THz) is generated via Difference-Frequency Generation (DFG), see level scheme in Fig.~\ref{Fig:setup}. The idler field is an amplified replica of the signal field.  The crystal and strong pump beam act as a phase-insensitive amplifier, meaning that the gain is independent of the phase of the input signal field. The ultrafast nature of this scheme stems from the ability to utilize a picosecond pulsed pump to achieve the conversion, allowing for the sampling of picosecond-scale temporal information from the signal field and transferring it into the idler field through stimulated parametric amplification.
However, in addition to the signal and idler fields, the pump spontaneously amplifies vacuum fluctuations at the wavelength of both signal and idler. As we will see, these spontaneous fluorescence photons will have an important effect in the measurement of photon distributions of the photon field.

\subsection{Experimental setup and imaging}

\begin{figure}[t]
	\includegraphics[width=0.48\textwidth]{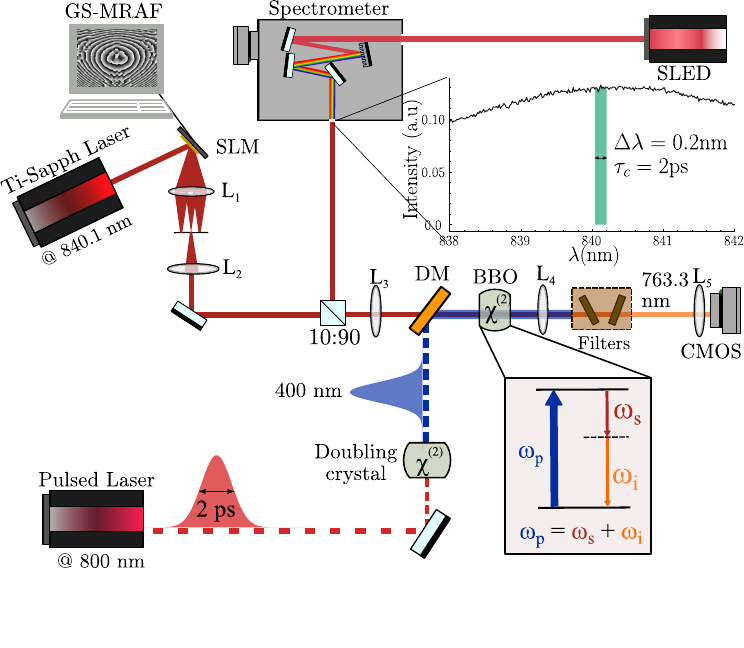}
	\caption{Experimental Setup. A 2~ps pump laser is sent to a BBO crystal where it is combined simultaneously with a coherent laser source (Ti:Sapph with typical power of 3-4 mW) and a thermal source (SLED with typical power of $40$ $\mu$W after spectral filtering). Down-converted photons are filtered and recorded using a CMOS camera. The repetition rate of the camera, 1KHz, matches that of the pulsed laser, allowing for single shot image recording.}
	\label{Fig:setup} 
\end{figure}

To demonstrate the capabilities of our set-up to measure photon statistics in space, we consider the case depicted in Fig.~\ref{Fig:setup} where the input state  sent into the system is constituted of two input fields with different statistical properties and spatial profiles. The first is a coherent state originated from a Ti:Sapph laser, which has been spatially tailored transversely using a Spatial Light Modulator (SLM) and a Gerchberg-Saxton-based Mixed Region Amplitude Freedom (GS-MRAF) algorithm to create a spatial image in the form of the letter "A" (See appendix for more details about the SLM). The second input field is generated from a thermal source  (Exalos, EXS210153-01) with a broad 20 nm spectrum, which we spectrally filter with a spectrometer to retain a spectrum of $0.2$ nm width centered at $840$ nm. This bandwidth ensures a coherence time of $2$ ps, suitable for observing the thermal source with picosecond resolution. Afterwards, we employ a beam-shaping method to refocus the thermal light into a small Gaussian beam with a waist of 15 microns. Both input fields are combined with the pulsed pump laser via a 10/90 beam splitter and imaged onto the BBO crystal.  After DFG amplification (type 1 $(o)+(o)\Rightarrow(e)$ with a crystal angle of 29.14) at the BBO crystal, a band-pass (FF01-747/33-25) and two high-pass and low-pass interference Semrock filters (TLP01-790.25x36 and TSP01-790-25x36) are added to select only the optical fields centered around 763.3 nm with roughly 1 nm bandwidth. The light emitted from the BBO crystal at these frequencies is then imaged on a Hamamatsu ORCA-Fusion CMOS camera. In this configuration, we collect 10000 single-shot 2D images at 1 kHz, the repetition frequency of the pulsed laser. 

\begin{figure}[t]
  \centering
  \includegraphics[width=0.48\textwidth]{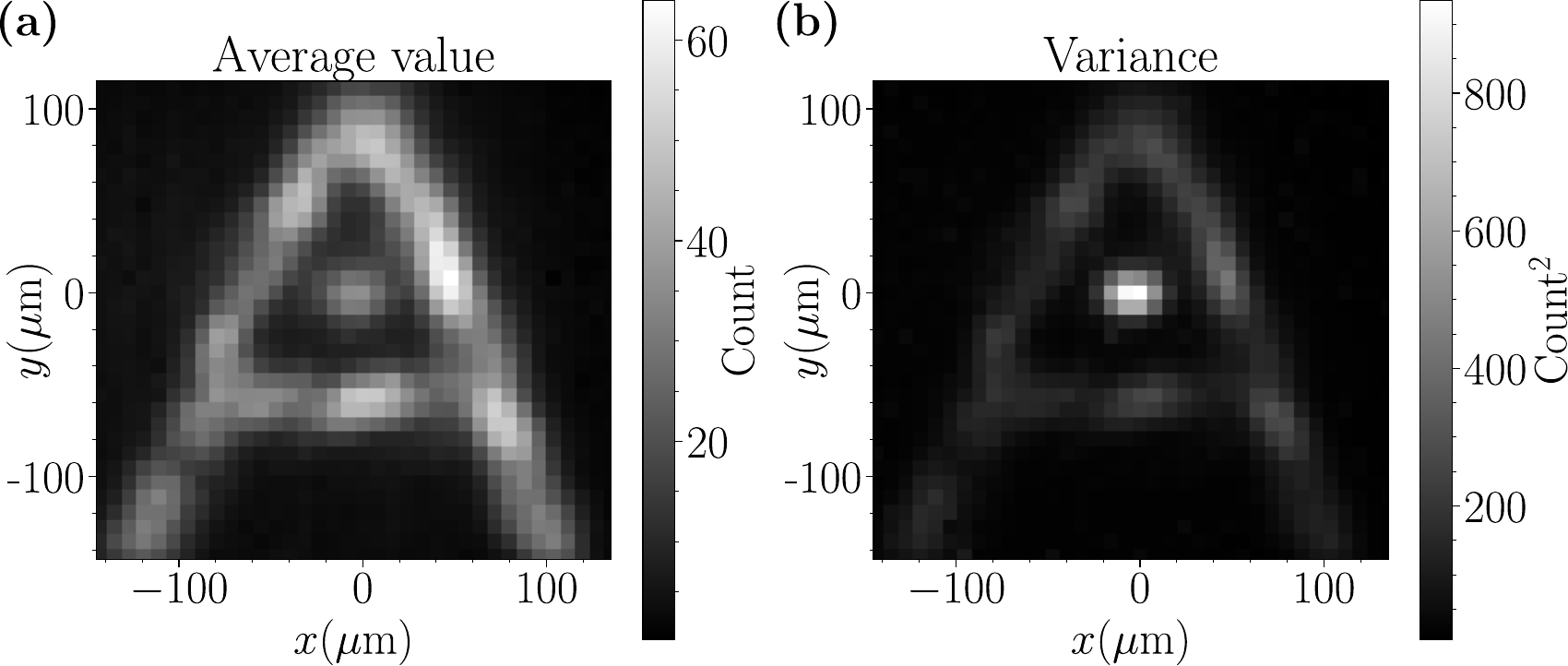}
    \caption{(a) Measured average number of photons in a two-dimensional image using 10000 single-shot. (b) Variance of the photon count over the image.}
  \label{fig:AmeanVar} 
\end{figure}

\begin{figure*}[t]
	\centering
	\includegraphics[width=1\linewidth]{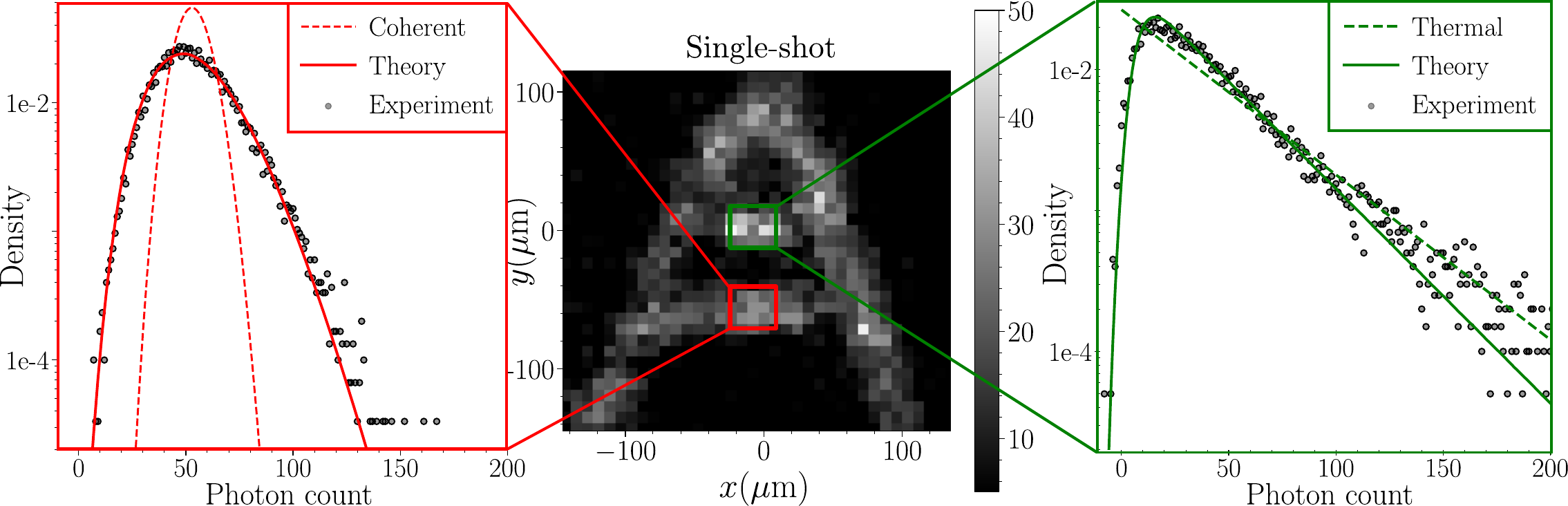}
	\caption{Center: Single shot of a two-dimensional spatial distribution and associated density distributions of the photon counts relative to two different positions in the image obtained from 10000 single-shot measurements. Dots are the experimental results, and in red and green lines are the theoretical curves calculated with $g_{\text{2D}}=5.963\times 10^{-13}$ s.m$^{-1}$ for both a coherent and a thermal state. The red and green dashed lines are coherent and thermal distribution, respectively, possessing the same average photon number as measured in the experiment.}
\label{Fig:DoubleHisto} 
\end{figure*}

Figure~\ref{fig:AmeanVar}(a) and (b) display, respectively, the average photon number and the variance on each pixel over the collection of 10000 images. A selection of the individual single shot images is shown in the appendix for completeness.
Firstly, we observe that the output spatial photon distribution clearly exhibits an A-shaped pattern, which corresponds to the input image generated by the SLM using the coherent continuous-wave laser source. This result is achieved because the phase-matching tolerance was carefully chosen, ensuring that the transverse $k$ components forming the image are all amplified with the same magnification. This measurement shows that the setup is capable of faithfully transforming complex images to another wavelength. 
Secondly, we observe a Gaussian shape spot at the center of the A-shaped distribution. 
This spot originates from the thermal light source and displays an average photon number similar to that of the coherent-source A shape. The variance at the small spot inside the A is much larger than that at the A shape itself. This important distinction indicates that the statistical distribution of photons forming the A shape is very different from that of the central small spot.

\subsection{2D photon distribution analysis}

We perform a quantitative analysis of the image by directly evaluating the probability of photon count number at two different positions within the image. To do so we use the 10000 single-shot images to construct two normalized histograms that define two photon distributions at two different positions marked in red and green squares in Fig.~\ref{Fig:DoubleHisto}. The two distributions are shown in semi-logarithmic plots.  The photon distribution measured at the green square location (right panel), originating from the thermal input field, shows after 30 photon counts per pixel, an exponential decay in photon number, a signature usually associated with thermal statistics. Nevertheless, if we compare with a pure thermal distribution that possesses the same average photon number as the experimental measurement (dashed green line)  we see a quantitative discrepancy for photon count between 0 and 50. This indicates that the measured distribution is not purely thermal. 

The left panel of  Fig.~\ref{Fig:DoubleHisto} displays the measured photon distribution from the red square, originating from the continuous wave laser source. We can compare the distribution with that of a coherent state with average photon number corresponding to the measured one (dashed red line). We see a strong disagreement between the two, indicating that the measured photon distribution is not purely the one associated with a coherent state. Despite these disagreements, we can still conclude that this two-dimensional detection setup allows us to qualitatively distinguish between the coherent and thermal photon statistics present in the input state. 

To be more quantitative, we can evaluate the intensity autocorrelation function

\begin{equation}\label{eq:g2}
g_{N}^{(2)}=\frac{\braket{N^{2}}}{\braket{N}^{2}}=1+\frac{\text{Var}(N)}{\braket{N}^{2}},
\end{equation}
in the different regions of the image. In Eq.~\eqref{eq:g2}, $N$ corresponds to the photon counts at a chosen location in the image and $\braket{N}$ its average value. 
However, as introduced above, the DFG set-up introduces amplified vacuum fluorescence at the wavelength of the measured idler field. 
This fluorescence also has an average photon number value $\braket{N_{\text{vac}}}$ and variance $\text{Var}(N_{\text{vac}})$ that we denote as ``vac'' for vacuum for clarity, and it can be directly measured in an experiment with no input signals.
To realize a precise computation of $g_{N}^{(2)}$ of the idler field, we need to remove the photon count contribution originating in the fluorescence from the measured variance and average photon count.
This can be directly done because photons emitted by amplified spontaneous processes constitute independent processes compared to photons emitted via stimulated parametric processes. In the upcoming sections we will treat this in detail.  Defining the corrected quantities:
\begin{equation}
\begin{split}
\braket{N_{\text{corr}}}&=\braket{N}-\braket{N_{\text{vac}}}, \\
\text{Var}(N_{\text{corr}})&=\text{Var}(N)-\text{Var}(N_{\text{vac}}),
\end{split}
\end{equation}
we deduce a corrected autocorrelation $g^{(2)}_{N_{\text{corr}}}$:
\begin{equation}
g^{(2)}_{N_{\text{corr}}}=1+\frac{\text{Var}(N_{\text{corr}})}{\braket{N_{\text{corr}}}^{2}}.
\end{equation}

Applying these formulas to the photon counts measured at the position of the green square in Fig.~\ref{Fig:DoubleHisto} we obtain $g^{(2),\text{thermal}}_{N_{\text{corr}}}=2.15$, which is close to the theoretical value of 2 expected for an ideal thermal source. Similarly, for photon counts measured at the position of the red square in Fig.~\ref{Fig:DoubleHisto} we obtain $g^{(2),\text{coherent}}_{N_{\text{corr}}}=1.31$, which is 30$\%$ off compared to the ideal value of 1 for a perfect coherent source. 

The two measurements of $g^{(2)}$ values and the full photon distributions at the green and red square locations differ both from the ideal thermal and coherent case. In the rest of this paper we will show that the difference between the measured distributions and those of the input signal are of fundamental origin. They arise from the amplification of spontaneous vacuum fluctuations and from the multimodal nature of the system ``nonlinear BBO crystal + strong pump'' acting as an amplifier. 
In the next sections, we propose to investigate quantitatively these aspects and show how we can describe comprehensibly the measured photon distributions using a multimodal approach. The multimodal approach allows us to have predictions that are plotted as solid lines in Fig.~\ref{Fig:DoubleHisto}. They capture remarkably well the measured distributions.

\section{Temporal multimodal Description}\label{sec:multimodal}

In this section we present analytical, numerical, and experimental results that introduce and explain the temporal multimodal description of the DFG amplifier. In the following parts, for pedagogical purposes and to increase the complexity of the description progressively, we restrict the analysis to a ``0D'' description where we assume that all input fields are plane waves. We expect this approximation to hold as long as the transverse wavevector $k$ components of the image, which are converted through the DFG process, are all amplified with the same gain owing to the phase-matching tolerance chosen for the BBO crystal. The experiments described in this section used a different BBO, compared to the case shown previously, with a size of $ l_c=3$ mm, optimized to work with $\lambda_s=853$ nm and $\lambda_i=753$ nm.

\subsection{Singular Value Decomposition, Bogoliubov transformation and input-ouput relation}
\label{SM:SVD_Bogoliubov}
This part introduces the mathematical tools and conventions to perform the modal treatment of the DFG process and follows references ~\cite{Wasilewski2006,Patera2012,Christ2013}.
We decompose the positive part of the electric field in terms of the Fourier components 
\begin{equation}
\resizebox{0.92\columnwidth}{!}{$
\hat{E}^{(+)}(t,z)=\sum_{n=\mathrm{s},\mathrm{i}}
\mathcal{E}_{n}
\int\mathrm{d}\Omega\;
\mathrm{e}^{-\mathrm{i}(\omega_n+\Omega) t}
\mathrm{e}^{\mathrm{i}k_z(\omega_n+\Omega)z}
\hat{a}_n(\Omega,z),
$}
\end{equation}
where $\Omega$ are sidebands with respect to the carrier $\omega_n$ that can take the value $n=\mathrm{i}$ for $\omega_\mathrm{i}$ and $n=\mathrm{s}$ for $\omega_\mathrm{s}$ and with normalization factor
\begin{equation}
\mathcal{E}_{n}=\sqrt{\frac{\hbar\omega_n^2}{2\epsilon_0 c^2 k_n}},
\end{equation}
where $\epsilon_0$ is the permittivity of vacuum, $c$ is the speed of light in vacuum, $\hbar$ is the reduced Planck constant, and $k_n$ the wave-vector of the optical field at frequency $\omega_n$. $\mathcal{E}_{n}$ has the meaning of a single photon field amplitude. The operators $\hat{a}_n^{\dagger}(\Omega,z)$ and $\hat{a}_n(\Omega,z)$ respectively create and annihilate photons at frequency $\Omega$ at position $z$  and obey the commutation relation
\begin{equation}
\left[ \hat{a}_n(\Omega,z), \hat{a}^{\dagger}_{n'}(\Omega',z') \right]=\delta_{n,n'}\delta(\Omega-\Omega')\delta(z-z').  
\end{equation}
$\delta(\Omega-\Omega')$ and $\delta(z-z')$ are Dirac delta functions and $\delta_{n,n'}$ is the Kronecker delta. The index $n=\{\mathrm{s},\mathrm{i}\}$ accounts for the signal and idler fields collinearly propagating along the $z$-axis through the nonlinear medium. The interactions between these fields are mediated by a strong classical pump pulse of spectral amplitude $A_{\mathrm{p}}$, and is such that we have perfect phase matching at carrier frequencies: $\omega_\mathrm{p}=\omega_\mathrm{s}+\omega_\mathrm{i}$ and $\bm{k}_\mathrm{p}=\bm{k}_\mathrm{s}+\bm{k}_\mathrm{i}$. The wavevectors $\bm{k}_\mathrm{p}$, $\bm{k}_\mathrm{s}$, and $\bm{k}_\mathrm{i}$ are associated to the pump, signal, and idler fields, respectively. In the undepleted pump approximation (the pump amplitude does not vary over the propagation through the crystal i.e., it does not depend on $z$), the evolution of signal and idler is given by~\cite{Christ2013}
\begin{align}
\frac{\partial}{\partial z}\hat{a}_\mathrm{s}(\Omega,z)&=
g\int\mathrm{d}\Omega'\;
K(\Omega,\Omega',z)\,
\hat{a}_\mathrm{i}^{\dag}(\Omega',z),
\label{evoas}
\\
\frac{\partial}{\partial z}\hat{a}_\mathrm{i}(\Omega,z)&=
g\int\mathrm{d}\Omega'\;
K(\Omega',\Omega,z)
\hat{a}_\mathrm{s}^{\dag}(\Omega',z),
\label{evoai}
\end{align}
where $g$ is a parameter that depends on the crystal and field properties (see Appendix for more detail), $K(\Omega,\Omega',z)=A_\mathrm{p}(\Omega+\Omega')\mathrm{e}^{\mathrm{i}\Delta(\Omega,\Omega') z}$ is the kernel of the two integro-differential equations, $\Delta(\Omega,\Omega')=k_{\mathrm{p}z}(\Omega+\Omega')-k_{\mathrm{s}z}(\Omega)-k_{\mathrm{i}z}(\Omega')$ is the phase-mismatch angle that takes into account for the dispersion and the birefringent character of the nonlinear crystal, and $k_{\mathrm{n}z}(\Omega)=k_{z}(\omega_n+\Omega)$ (for $n\in\{\mathrm{s},\mathrm{i}\}$) is the $z$ component of the signal and idler wave-vectors. $A_\mathrm{p}(\Omega+\Omega')$ is the unitless spectrum amplitude of the pump pulse (see Appendix for more details).

Since Eqs.~\eqref{evoas} and~\eqref{evoai} are linear in the field operators, a formal solution can be obtained in terms of a linear integral transformation (see for example~\cite{Wasilewski2006,Patera2012,Christ2013}), generalizing the usual Bogoliubov transformation. Since $K$ is $z$-dependent, the exact solution can only be obtained by numerical integration. On the other hand, it has been shown~\cite{Christ2013} that the Magnus expansion truncated at first order provides an accurate analytical description up to a degree of amplification of about $12$dB in fields, which corresponds to an amplification in photon number of a factor 16. In other words, this approximation is valid as long as the system gain is smaller than 16. The advantage of using truncated Magnus expansion, as we will show below, is that the solution can be cast in explicit form in terms of the normal modes of the evolution, offering a visualization and understanding of the modes in which photons are emitted. These modes are also known as ``supermodes'' or ``temporal'' modes; we will use the denomination ``temporal modes'' in this paper.

We start by rewriting Eqs.~\eqref{evoas} and~\eqref{evoai} in compact form, by defining:
\begin{equation}
\hat{\bm{\xi}}(\Omega,z)=
\left(
\begin{array}{c}
\hat{a}_{\mathrm{s}}(\Omega,z)
\\
\hat{a}_{\mathrm{i}}(\Omega,z)
\\
\hat{a}_{\mathrm{s}}^{\dagger}(\Omega,z)
\\
\hat{a}_{\mathrm{i}}^{\dagger}(\Omega,z)
\end{array}
\right),
\end{equation}
so that the field evolution equation becomes
\begin{equation}
\frac{\partial \hat{\bm{\xi}}(\Omega,z)}{\partial z}=
g \int\mathrm{d}\Omega' \,
\mathbb{K}(\Omega,\Omega',z)\,
\hat{\bm{\xi}}(\Omega',z),
\label{eq:evo}
\end{equation}
with
\begin{equation}
\mathbb{K}(\Omega,\Omega',z)=
\resizebox{0.7\columnwidth}{!}{$
\left(
\begin{array}{cccc}
0 & 0 & 0 & K(\Omega,\Omega',z)
\\
0 & 0 & K(\Omega',\Omega,z) & 0
\\
0 & K^*(\Omega,\Omega',z) & 0 & 0
\\
K^*(\Omega',\Omega,z) & 0 & 0 & 0
\end{array}
\right),
$}
\label{Bogo kernel}
\end{equation}
As mentioned above, the solution of the $z$-dependent problem, Eq.~\eqref{eq:evo}, can be obtained in terms of the Magnus expansion truncated at the first order. This is equivalent to the infinite Dyson series when $z$-ordering is neglected.  Then, associating an ``out'' label to the position after the crystal size at $z=l_c/2$ ($\hat{\bm{\xi}}_{\mathrm{out}}(\Omega)\coloneqq\hat{\bm{\xi}}(\Omega,l_c/2)$) and a ``in'' label for the position just before the crystal at $z=-l_c/2$ ($\hat{\bm{\xi}}_{\mathrm{in}}(\Omega)\coloneqq\hat{\bm{\xi}}(\Omega,-l_c/2)$), we do the Magnus expansion and integrate over the crystal size by considering the following linear and symplectic transformation:
\begin{equation}
\hat{\bm{\xi}}_{\mathrm{out}}(\Omega)=\int\mathrm{d}\Omega'\,\mathrm{Exp}\left[g l_c\mathbb{M}_1(\Omega,\Omega')\right]\hat{\bm{\xi}}_{\mathrm{in}}(\Omega'),
\label{Exp solution}
\end{equation}
with 
\begin{equation}
\mathbb{M}_1(\Omega,\Omega')=
\resizebox{0.7\columnwidth}{!}{$
\left(
\begin{array}{cccc}
0 & 0 & 0 & \widetilde{M}_1(\Omega,\Omega')
\\
0 & 0 & \widetilde{M}_1(\Omega',\Omega) & 0
\\
0 & \widetilde{M}_1^*(\Omega,\Omega') & 0 & 0
\\
\widetilde{M}_1^*(\Omega',\Omega) & 0 & 0 & 0
\end{array}
\right),
$}
\label{Bogo kernel}
\end{equation}
 and 
\begin{align}
 \nonumber
\widetilde{M}_1(\Omega,\Omega')&=
\int_{-l_{c}/2}^{l_{c}/2}\,K^{*}(\Omega,\Omega',z)\mathrm{d}z
\\ 
&=A^{*}_\mathrm{p}(\Omega+\Omega')\,\mathrm{Sinc}\left(\Delta(\Omega,\Omega')\frac{l_c}{2}\right).
\label{eq:JSA}
\end{align}

$\widetilde{M}_1(\Omega,\Omega')$  represents the generic Joint Spectral Amplitude (JSA) that describes the coupling between sidebands at frequency $\Omega$ and sidebands at $\Omega'$, and $\mathrm{Sinc(x)}=\sin(x)/x$ is the sinc function. 

From the structure of Eq.~\eqref{eq:JSA}, it is clear that the basis of Fourier modes originally employed for describing the electric fields is not suitable for understanding the results of the dynamics of the pulsed DFG: i.e., plane-waves are not the eigenmodes of the problem. Instead, it is desirable to find the normal modes that decouple the solution of Eq.~\eqref{Exp solution}. This can be obtained by performing the Singular Value Decomposition (SVD) of the JSA (Eq.~\eqref{eq:JSA}) which leads to:
\begin{equation}
\widetilde{M}_1(\Omega,\Omega')=\sum_m \lambda_m \psi_m(\Omega)\phi_m^*(\Omega'),
\label{M1 svd}
\end{equation}
where $\psi_m(\Omega)$ and $\phi_m(\Omega')$ are the temporal eigenmodes of the signal and the idler, respectively, and $\lambda_m\geq0$ is the corresponding singular value. It is interesting to note that this decomposition also induces a SVD of Eq.~\eqref{Exp solution} that preserves the symplectic structure of the problem, and it is known as Bloch-Messiah decomposition~\cite{Bennink2002,Law2000,deValcrcel2006,Wasilewski2006,Patera2012,Christ2013}. The latter can be written as 
\begin{equation}
\mathrm{Exp}\left[g l_c\mathbb{M}_1(\Omega,\Omega')\right]
=
\sum_m
U_m(\Omega) D_m U_m^\dagger(\Omega'),
\label{Eq:BlochMessiah}
\end{equation}
where
\begin{equation}
U_m(\Omega)=
\left(
\begin{array}{cccc}
\psi_m(\Omega) & 0 & 0 & 0 \\
0 & \phi_m^*(\Omega) & 0 & 0 \\
0 & 0 & \psi_m^*(\Omega) & 0 \\
0 & 0 & 0 & \phi_m(\Omega)
\end{array}
\right),
\end{equation}
and

\begin{equation}
D_m =
\resizebox{0.8\columnwidth}{!}{$
\left(
\begin{array}{cccc}
\cosh(g\lambda_m l_c) & 0 & 0 & \sinh(g\lambda_m l_c) \\
0 & \cosh(g\lambda_m l_c) & \sinh(g\lambda_m l_c) & 0 \\
0 & \sinh(g\lambda_m l_c) & \cosh(g\lambda_m l_c) & 0 \\
\sinh(g\lambda_m l_c) & 0 & 0 & \cosh(g\lambda_m l_c)
\end{array}
\right).
$}
\label{Eq:CouplingMatrix}
\end{equation}

The two families of orthonormal functions $\{\psi_m(\Omega)\}_{m\in\mathbb{N}}$ and $\{\phi_m(\Omega)\}_{m\in\mathbb{N}}$ represent the components of the normal modes of the DFG in the Fourier basis. Thus, the normal modes in the temporal domain are given by
\begin{align}
u_{\mathrm{s},m}(z,t)&=
\int\mathrm{d}\Omega\,
\psi_m(\Omega)
\mathrm{e}^{-\mathrm{i}(\omega_{\mathrm{s}}+\Omega) t}
\mathrm{e}^{\mathrm{i}k_z(\omega_{\mathrm{s}}+\Omega)z},
\\
v_{\mathrm{i},m}(z,t)&=
\int\mathrm{d}\Omega\,
\phi_m(\Omega)
\mathrm{e}^{-\mathrm{i}(\omega_{\mathrm{i}}+\Omega) t}
\mathrm{e}^{\mathrm{i}k_z(\omega_{\mathrm{i}}+\Omega)z},
\end{align}
with the wavevector, $k_z$, having an explicit frequency dependence depending on the properties of the nonlinear medium. Most importantly, we can thus write explicitly the field (input/output) amplitude operators of these modes as
\begin{align}
\hat{c}^{m}_{\mathrm{s}}&=
\int\mathrm{d}\Omega\,
\psi_m(\Omega)\hat{a}_{\mathrm{s},\mathrm{in}}(\Omega),\\
\hat{c}^{m}_{\mathrm{i}}&=
\int\mathrm{d}\Omega\,
\phi_m(\Omega)\hat{a}_{\mathrm{i},\mathrm{in}}(\Omega),\\
\hat{b}^{m}_{\mathrm{s}}&=
\int\mathrm{d}\Omega\,
\psi_m(\Omega)\hat{a}_{\mathrm{s},\mathrm{out}}(\Omega),\\
\hat{b}^{m}_{\mathrm{i}}&=
\int\mathrm{d}\Omega\,
\phi_m(\Omega)\hat{a}_{\mathrm{i},\mathrm{out}}(\Omega),
\end{align}
where we use the letter $c$ for the input and the $b$ for the output.
In this basis, the problem is written as a set of decoupled two-mode Bogoliubov transformations
\begin{equation}
\left(
\begin{array}{c}
\hat{b}^{m}_{\mathrm{s}}
\\
\hat{b}^{m}_{\mathrm{i}}
\\
\hat{b}^{m,\dagger}_{\mathrm{s}}
\\
\hat{b}_{\mathrm{i}}^{m\dagger}
\end{array}
\right)
=D_m
\left(
\begin{array}{c}
\hat{c}_{\mathrm{s}}^{m}
\\
\hat{c}_{\mathrm{i}}^{m}
\\
\hat{c}_{\mathrm{s}}^{m\dagger}
\\
\hat{c}_{\mathrm{i}}^{m\dagger}
\end{array}
\right).
\label{Exp solution decoupled}
\end{equation}

\begin{figure*}[t!]
	\centering
	\includegraphics[width=0.80\textwidth]{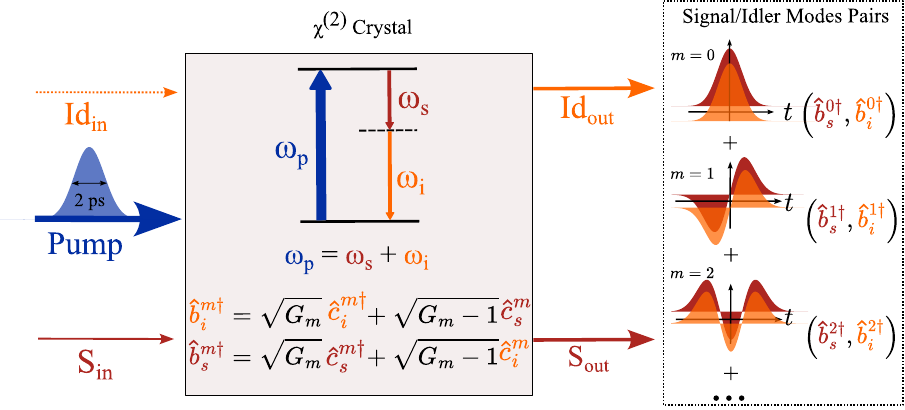}
	\caption{Schematic representation of the DFG process in the temporal eigenmode basis. }
	\label{Fig:tmodes} 
\end{figure*}

We can now focus on the possibilities of creating idler photons at the output of the crystal. From Eq.~\eqref{Exp solution decoupled}, we can identify two different independent processes. Firstly, we see that the lower left term of $D_m$ (Eq.~\eqref{Eq:CouplingMatrix}) allows for the conversion of an input signal photon in a given mode $m$ into an output idler photon in the same mode. Secondly, the lower right term is responsible for, in a first place, stimulated emission of an idler field in the mode $m$ at the output due to an existing idler field in the mode $m$ at the input, and in the second place, spontaneous amplification of an idler vacuum field in a given mode $m$ at the output. 
Thus, we can summarize these different processes with the following  input-output relation for each crystal eigenmode:
 \begin{equation}
\hat{b}^{m\dagger}_{\mathrm{i}}=\sinh{(g\lambda_m l_c)}\hat{c}^{m}_{\mathrm{s}}+\cosh{(g\lambda_m l_c)} \hat{c}_{\mathrm{i}}^{m\dagger}.
\label{Eq:InputOupoutSh}
\end{equation}

Defining the mode gain as
\begin{equation}
G_{m}\coloneqq\cosh^{2}{(g \lambda_m l_c)},
\label{eq:Gain_lambda_link}
\end{equation}
we can rewrite Eq.~\eqref{Eq:InputOupoutSh} as
\begin{equation}
\hat{b}_{\mathrm{i}}^{m\dagger}=\sqrt{G_m-1}\hat{c}^{m}_{\mathrm{s}}+\sqrt{G_m} \hat{c}_{\mathrm{i}}^{m\dagger}.
\label{Eq:InputOupoutSh1}
\end{equation}

For completeness, we also give the relation of the produced signal photons that can be obtained in the same manner:
\begin{equation}
\hat{b}_{\mathrm{s}}^{m\dagger}=\sqrt{G_m-1}\hat{c}^{m}_{\mathrm{i}}+\sqrt{G_m} \hat{c}_{\mathrm{s}}^{m\dagger}.
\label{Eq:InputOupoutSh2}
\end{equation}

We have derived the fundamental equations that link the photon generated at the output of the crystal with respect to photons present at the input. As a result of the SVD, each mode $m$ is independent and follows the same equation with a gain value $G_m$ that depends on the singular value $\lambda_m$ of the decomposition.  A schematic representation of this process is given in Fig. \ref{Fig:tmodes}. An input signal $\text{S}_{\text{in}}$ and an idler $\text{Id}_{\text{in}}$ field are decomposed onto a set of orthonormal input modes $\{\psi_m,\phi_m\}$. 
Each mode pair $m$ is independently amplified with a gain $G_m$. The total output fields ($\text{S}_{\text{out}}$ and $\text{Id}_{\text{out}}$) are reconstructed from these amplified components. The dashed line in the input Idler field in Fig.~\ref{Fig:tmodes} represents the process in which a vacuum idler field gets amplified to produce a nonzero average photon number at the output. These photons are usually referred to as fluorescence photons, which were already mentioned at the beginning of this paper.

\subsection{Analytical and numerical structure of the temporal modes}
\label{SM:NumAnaTemporalmodes}
In the following, we aim to obtain the spectral form of the set of modes $\psi_m$ and $\phi_m$ and the singular values $\lambda_m$. 
Experimentally, we only measure the photons emitted at around $\lambda_{\text{i}}= 750$ nm, and signal photons emitted at around $\lambda_{\text{s}}=840$ nm are filtered out via optical filters. 
We thus restrict the analysis on the emitted idler photons. Treating the emitted signal photons presents no difficulties. To make the connection between the experimental measurements and the theoretical description above explicit, we identify specific idler and signal sidebands in the Joint Spectral Amplitude (JSA) and define
\begin{align}
M_1(\Omega_\mathrm{i},\Omega_\mathrm{s}) \coloneqq \widetilde{M}^{*}_1(\Omega'=\Omega_\mathrm{i},\Omega=\Omega_\mathrm{s})\\ \nonumber= 
A_\mathrm{p}(\Omega_\mathrm{i}+\Omega_\mathrm{s})\,\mathrm{Sinc}\,\left(\Delta(\Omega_\mathrm{i},\Omega_\mathrm{s})\frac{l_c}{2}\right),
\label{eq:JSAGood}
\end{align}
where $\Omega_\text{s}=\omega-\omega_\text{s}$ and $\Omega_\text{i}=\omega-\omega_\text{i}$. The phase-mismatch, responsible for the frequency-dependent conversion efficiency in the BBO crystal, is generically expressed as
$\Delta(\Omega_\mathrm{i},\Omega_\mathrm{s})=k_{\mathrm{p}z}(\Omega_\mathrm{i}+\Omega_\mathrm{s})
-k_{\mathrm{s}z}(\Omega_\mathrm{i})-k_{\mathrm{i}z}(\Omega_\mathrm{s})$.
In the plane-wave approximation, where the three fields only possess a $k$-component along $z$ (also called co-linear phase-matching), $\Delta(\Omega_i,\Omega_s)$ can be approximated in terms of a Taylor expansion truncated to the second order,
\begin{equation}
\begin{aligned}
\Delta(\Omega_\mathrm{i},\Omega_\mathrm{s})\approx (k_{\mathrm{p}}-k_{\mathrm{s}}-k_{\mathrm{i}})+  \\  \left[k'_{\mathrm{p}}(\Omega_i+\Omega_s) k'_{\mathrm{s}}\Omega_\mathrm{s}-k'_{\mathrm{i}}\Omega_\mathrm{i}\right] +\\ 
\frac{1}{2}\left[k''_{\mathrm{p}}(\Omega_\mathrm{i}+\Omega_\mathrm{s})^2-k''_{\mathrm{s}}\Omega_\mathrm{s}^2-k''_{\mathrm{i}}\Omega_\mathrm{i}^2\right].
\label{Eq:mismatch}
\end{aligned}
\end{equation}
In this expression, $k_{\mathrm{p}}$, $k_{\mathrm{s}}$ and $k_{\mathrm{i}}$ represent the wave-vectors at the carrier of pump, signal, and idler respectively. Similarly,
$k'_{\mathrm{p}}$, $k'_{\mathrm{s}}$ and $k'_{\mathrm{i}}$ represent the inverse of the group velocity $dk/d\omega$ at the pump, signal, and idler carrier,
and $k''_{\mathrm{p}}$, $k''_{\mathrm{s}}$ and $k''_{\mathrm{i}}$ represent the second order dispersion $d^2k/d\omega^2$ at the pump, signal, and idler carrier. We consider the experimental situation where the three waves are phase-matched ($k_{\mathrm{p}}-k_{\mathrm{s}}-k_{\mathrm{i}}=0$) according to the Type I configuration in which the pump at $\lambda_{\mathrm{p}}=400$ nm with polarization along the extraordinary axis produces a signal at $\lambda_{\mathrm{s}}=853$ nm and an idler at $\lambda_{\mathrm{i}}=753$ nm, both polarized along the ordinary axis. The frequency-dependent refraction index for a BBO crystal is given by the Sellmeier equations,
\begin{equation}
n(\lambda)=\sqrt{A+\frac{B}{\lambda^2-C}-D\lambda^2}.
\end{equation}

\begin{figure}[t]
  \centering
  \includegraphics[width=0.48\textwidth]{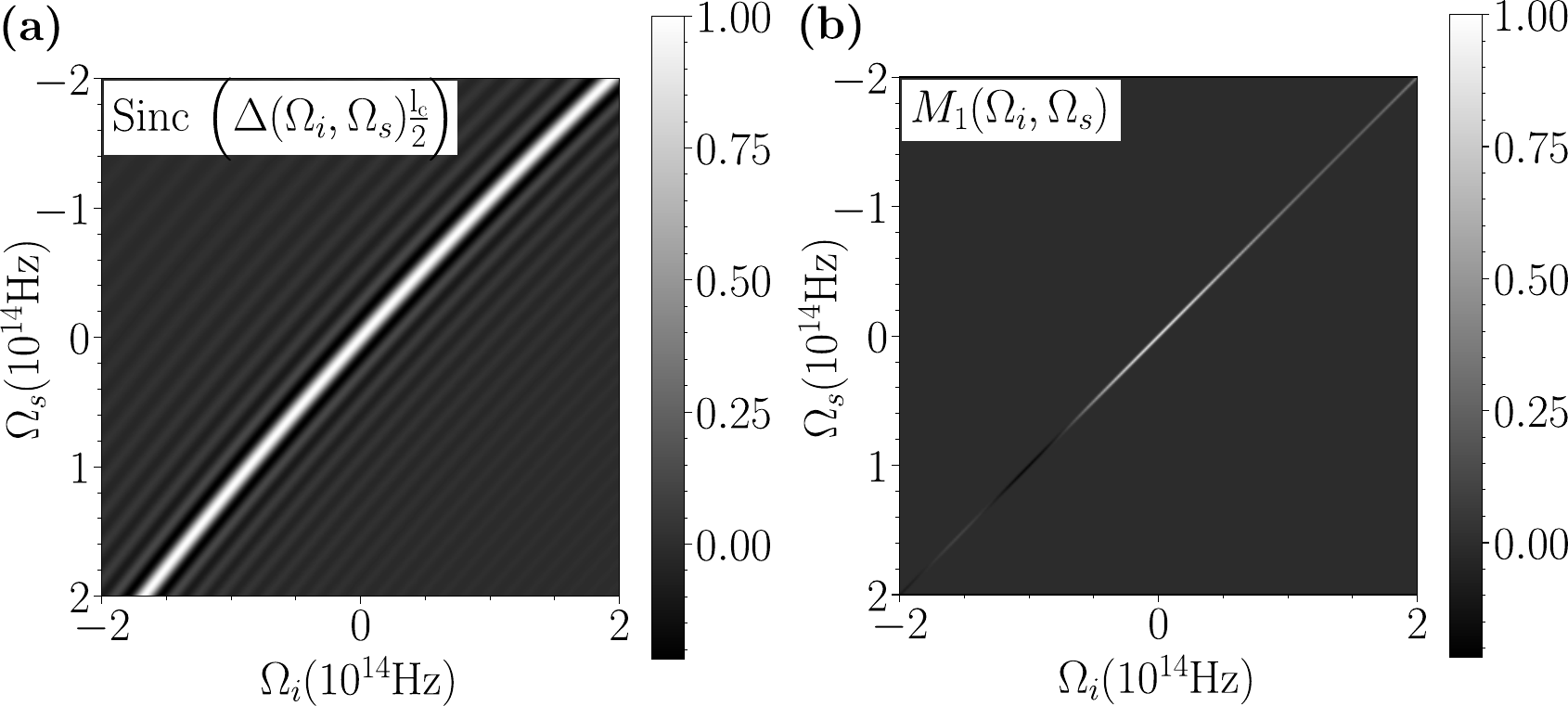}
    \caption{(a) Phase-matching and (b) the Joint Spectra Amplitude (JSA)  profile in the $(\Omega_i,\Omega_s)$ space.}
    \label{fig:PM_M1matrix}
\end{figure}

A BBO crystal is an uni-axial nonlinear medium and, in its reference frame, the x and y axis are characterized by
an ordinary refraction index $n_o(\lambda)$ with Sellmeier coefficients $A_o=2.7359$, $B_o=0.018782$ $\mu $m$^{2}$, $C_o=0.01822$ $\mu $m$^{2}$ and $D_o=0.01354$ $\mu $m$^{-2}$, 
while the z-axis is characterized by an extraordinary refraction index $n_e(\lambda)$ with Sellmeier coefficients $A_e=2.3753$, $B_e=0.01224$ $\mu $m$^{2}$, $C_e=0.01667$ $\mu $m$^{2}$ and $D_e=0.01516$ $\mu $m$^{-2}$. For the three selected wavelengths, the perfect phase-matching condition ($k_{\mathrm{p}}-k_{\mathrm{s}}-k_{\mathrm{i}}=0$) is found when the corresponding waves propagate
at an angle $\theta=29.01$ deg with respect to the optical axis of the crystal. At this angle, the extraordinary refraction index seen by the pump is given by
\begin{equation}
n_{\mathrm{p}}=\left(\frac{\sin^2(\theta)}{n_{e}(\lambda_{\mathrm{p}})^2}+\frac{\cos^2(\theta)}{n_{o}(\lambda_{\mathrm{p}})^2}\right)^{-1/2}.
\end{equation}
On the other hand, the signal and idler waves see the ordinary refraction index $n_o$ evaluated at their wavelengths. Knowing the dispersion parameters, see Table~\ref{tab:dispersion}, one can evaluate Eq.~\eqref{Eq:mismatch} and plot the phase-mismatch profile in Fig.~\ref{fig:PM_M1matrix}(a) for a crystal length of $l_\mathrm{c}=3$ mm.

\begin{table}[h!]
\centering
\footnotesize
\begin{tabular}{l|c}
\hline\hline
 & $\lambda_{\mathrm{p}}=400\operatorname{nm}$, $\lambda_{\mathrm{s}}=853\operatorname{nm}$, $\lambda_{\mathrm{i}}=753\operatorname{nm}$ \\
\hline
$k'_{\mathrm{p}}\,(\times10^{-9}\operatorname{s\,m^{-1}})$ & $5.81385$ \\
$k'_{\mathrm{s}}\,(\times10^{-9}\operatorname{s\,m^{-1}})$ & $5.60459$ \\
$k'_{\mathrm{i}}\,(\times10^{-9}\operatorname{s\,m^{-1}})$ & $5.62645$ \\
\hline
$k''_{\mathrm{p}}\,(\times10^{-25}\operatorname{s^{2}\,m^{-1}})$ & $1.96116$ \\
$k''_{\mathrm{s}}\,(\times10^{-26}\operatorname{s^{2}\,m^{-1}})$ & $6.71907$ \\
$k''_{\mathrm{i}}\,(\times10^{-26}\operatorname{s^{2}\,m^{-1}})$ & $8.20217$ \\
\hline
\end{tabular}
\caption{Dispersion parameters for critical phase-matching of BBO at room temperature.}
\label{tab:dispersion}
\end{table}

Multiplying the pump spectral function and the phase-matching functions gives the joint spectral amplitude $M_1(\Omega_i,\Omega_s)$ introduced earlier:
\begin{equation}
M_1(\Omega_i,\Omega_s)=
A_\mathrm{p}(\Omega_\mathrm{i}+\Omega_\mathrm{s})\,\mathrm{Sinc}\,\left(\Delta(\Omega_\mathrm{i},\Omega_\mathrm{s})\frac{l_\mathrm{c}}{2}\right),
\label{eq:JSA_sm}
\end{equation}
which is plotted in Fig.~\ref{fig:PM_M1matrix}(b).
It shows the strength of the frequency coupling in the nonlinear crystal, accounting for the spectral limitation of the pump. The pump's characterization is provided in the appendix. As explained in the previous part, temporal modes can generally be obtained through the numerical SVD of the JSA. However, apart from configurations where the pump and phase-matching are specifically engineered to tailor quantum states~\cite{Patera2012,Serino2024}, the JSA in most experimental situations can be well approximated by a bi-variate that we define as $M_1^{\text{ana}}(\Omega_\mathrm{i},\Omega_\mathrm{s})$:
\begin{align}
M_1(\Omega_\mathrm{i},\Omega_\mathrm{s})\approx M^{\text{ana}}_1(\Omega_\mathrm{i},\Omega_\mathrm{s})   \nonumber \\ 
\coloneqq\mathrm{e}^{-(\Omega_\mathrm{i}+\Omega_\mathrm{s})^2/2\Delta_1^2}\,
\mathrm{e}^{-(\Omega_\mathrm{i}-\Omega_\mathrm{s})^2/2\Delta_2^2},
\label{Eq:M1ana}
\end{align}
where $\Delta_1$ accounts for the JSA bandwidth along the direction $\Omega_\mathrm{i}-\Omega_\mathrm{s}=0$ and $\Delta_2$ accounts for the JSA bandwidth along the direction $\Omega_\mathrm{i}+\Omega_\mathrm{s}=0$. This Gaussian approximation is valid as long as the frequency window of the emitted photons is small compared to the bending of the phase-matching due to second-order dispersion terms.
\\

By using the approximation $\mathrm{Sinc}(x)\approx \mathrm{exp}(-x^2/5)$, aiming at matching the FWHM of the Sinc function by an exponential, we identify Eq.~\eqref{Eq:M1ana} to Eq.~\eqref{eq:JSA_sm} to obtain $\Delta_1^{-2}=8\ln(2)\Delta_\omega^{-2}+\Delta_{\mathrm{PM}}^{-2}$ with $\Delta_{\mathrm{PM}}^{-2}\approx (2k'_{\mathrm{p}}-k'_{\mathrm{s}}-k'_{\mathrm{i}})^2 l_c^2 / 40$. Since $\Delta_{\mathrm{\omega}}\approx1.38\times10^{12}$sec$^{-1}$ and $\Delta_{\mathrm{PM}}\approx5.31\times10^{12}$sec$^{-1}$, then $\Delta_1\approx0.584\times10^{12}$sec$^{-1}$; $\Delta_2^{-2}=(k'_{\mathrm{i}}-k'_{\mathrm{s}})^2 l_c^2 / 40$, thus $\Delta_2\approx96.44\times10^{12}$sec$^{-1}$. 
These numbers correspond to the case of the experimental DFG with parameters shown in Table \ref{tab:dispersion} and with the pump spectrum characterization presented in the appendix. Thus, replacing $M_1(\Omega_i,\Omega_s)$ by $M^{\text{ana}}_1(\Omega_i,\Omega_s)$ in Eq.~\eqref{M1 svd} allows us to solve analytically the problem and obtain explicit solutions for the temporal modes in the form of Hermite-Gauss functions and an expression for the singular value in terms of a geometric progression. The obtained analytical results are
\begin{align}
\lambda^{\text{ana}}_m&=
\frac{\sqrt{2\pi}\Delta_1\Delta_2}{\Delta_1+\Delta_2}
\left|\frac{\Delta_1-\Delta_2}{\Delta_1+\Delta_2}\right|^{m},
\\
\psi^{\text{ana}}_m(\Omega_i)&=
\frac{2^{1/4}\tau^{1/2}}{\sqrt{m! 2^m \pi^{1/2}}}
\mathrm{e}^{-\tau^2 \Omega_i^2} H_m(\sqrt{2}\tau\Omega_i),
\\
\phi^{\text{ana}}_m(\Omega_s)&=\frac{2^{1/4}\tau^{1/2}}{\sqrt{m! 2^m \pi^{1/2}}}
\mathrm{e}^{-\tau^2 \Omega_s^2} H_m(\sqrt{2}\tau\Omega_s),
\label{Eq:HermiteModes}
\end{align}
where $\tau=1/\sqrt{\Delta_1\Delta_2}$ and $H_n(x)$ are Hermite polynomials.
\\

\begin{figure}[t]
  \centering
  \includegraphics[width=0.48\textwidth]{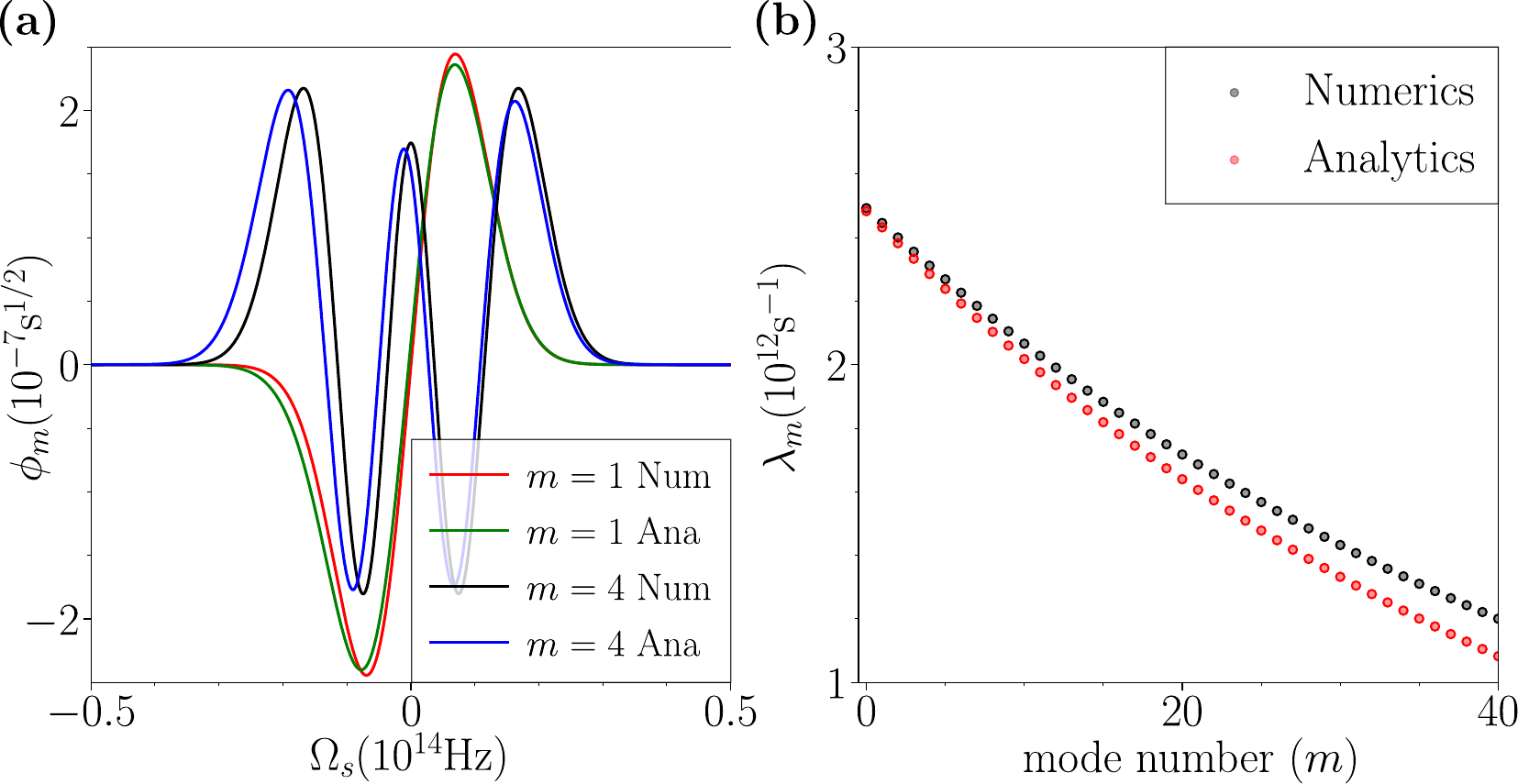}
    \caption{(a) Signal temporal eigenmodes $\phi_m(\Omega_\mathrm{s})$ for mode $m=4$ and $m=1$ obtained analytically (in red and blue) using the formula of Eq. (\ref{Eq:HermiteModes}) and obtained by performing the numerical SVD of the JSA (black and green). (b) Analytical and numerical singular value $\lambda_m$ associated with each mode $m$.}
    \label{fig:Spectrum_EGVAnaNumeric}
\end{figure}

To illustrate the validity of this description, we compare the analytical modes calculated using the Gaussian approximation and the numerical ones obtained through numerical SVD of the JSA. Figure \ref{fig:Spectrum_EGVAnaNumeric}(a) shows representative temporal modes in the spectral domain. The analytical ones are Hermite polynomials and thus even or odd. The numerical ones present a slight asymmetry, as the Kernel is not symmetric due to the second-order dispersion in the phase-matching term. However, the analytical formula describes well the temporal modes and allows us to understand their oscillatory behavior originating from the Hermite polynomial functions. It is important to note that the representation in the time domain of such modes is also close to a Hermite polynomial, and thus strongly oscillating. The latter are represented on the right side of Fig.~\ref{Fig:tmodes}. Finally, Fig.~\ref{fig:Spectrum_EGVAnaNumeric}(b) compares the numerical and the analytical singular values associated with each of the temporal modes. Again, there is an excellent agreement between the two, and the slight deviation observed is also due to the presence of second-order dispersion in the numerical calculation.

Although the analytical and numerical calculations agree well, we use the numerical results throughout this paper. The numerical approach incorporates second-order dispersion, which is present in the experiment, and is therefore used to derive key quantities such as photon statistics. The purpose of introducing the analytical modes here is to provide insights into the properties of the temporal modes and to understand their oscillatory nature in time and in the spectrum.

We have seen previously that the singular value decomposition of the joint spectral amplitude $M_1(\Omega_\mathrm{i},\Omega_\mathrm{s})$ can be written in the following form:
\begin{equation}
M_1(\Omega_\mathrm{i},\Omega_\mathrm{s})=\sum_m \lambda_m \psi_m(\Omega_\mathrm{i})\phi_m^*(\Omega_\mathrm{s}),
\label{eq:SM_M1_svd}
\end{equation}
where $\psi_m(\Omega_\mathrm{i})$ and $\phi_m(\Omega_\mathrm{s})$ are the respective crystal temporal eigenmodes of the idler and the signal associated with a common singular value $\lambda_m$. We recall that the $\psi_m(\Omega_\mathrm{i})$ are orthogonal to each other; the same applies for $\phi_m(\Omega_\mathrm{s})$. These modes are of fundamental importance because they constitute the basis where two photons, simultaneously emitted in the same set of modes $\phi_m$ and $\psi_m$ through spontaneous 
or stimulated parametric emission, are fully correlated (entangled) and are statistically independent from photons emitted in modes with different $m$. This allows us to understand why this temporal mode basis is naturally suited to describe problems associated to photon statistics.

\subsection{Input-Output formalism and vacuum statistics}
\label{SM:VacuumStat}
The evaluation of the photon number produced through the DFG process can be done considering the independent sets of eigenmodes obtained after singular value decomposition. From each set of eigenmode $m$ (for the idler and the signal) we have seen in section \ref{SM:SVD_Bogoliubov}, Eqs.\eqref{Eq:InputOupoutSh1} and  \eqref{Eq:InputOupoutSh2}, that we can write a set of coupled equations in an input-output formalism,
\begin{equation}
\begin{split}
b_\mathrm{i}^{m} &= u_{m} c_\mathrm{i}^{m}+ v_{m} {c_\mathrm{s}^{m}}^\dag,\\
b_\mathrm{s}^{m} &= u_{m} c_\mathrm{s}^{m}+ v_{m} {c_\mathrm{i}^{m}}^\dag,\\
\label{eq:SM_Input_Output}
\end{split}
\end{equation}
where $u_m=\sqrt{G_m}$ and $v_m=\sqrt{G_m-1}$ are related to the mode gain of Eq.~\eqref{eq:Gain_lambda_link}. 
\\

From Eq.~\eqref{eq:SM_Input_Output}, it is possible to compute the probability of having $n$ photons in the idler mode $m$ after the crystal (experimental situation of interest), starting from the vacuum.
For a given $m$, we can call $|0\rangle_{\rm in}$ the vacuum of the input modes (idler and signal), such that $c^m_\mathrm{s}|0\rangle_{\rm in}=c^m_\mathrm{i}|0\rangle_{\rm in}=0$. It is convenient to introduce the state $|0\rangle_{\rm out}$, the vacuum of the $b$ operators, $b^m_\mathrm{i}|0\rangle_{\rm out}=b^m_\mathrm{s}|0\rangle_{\rm out}=0$. Following \cite{Takagi1986}, one easily finds that
\begin{equation}
|0\rangle_{\rm in} = \frac{1}{u_m} e^{\frac{v_m}{u_m} {b^m_\mathrm{i}}^\dag {b^m_\mathrm{s}}^\dag  }|0\rangle_{\rm out}= \frac{1}{u_m}\sum_{n=0}^\infty\left(\frac{v_m}{u_m}\right)^n |n,n\rangle_{\rm out},
\end{equation}
where $|n,n'\rangle_{\rm out}$ corresponds to the state with $n$ photons of $b^m_\mathrm{i}$ and $n'$ of $b^m_\mathrm{s}$.

Therefore, the probability of finding exactly $n$ photons of $b^m_\mathrm{i}$ starting from the vacuum of $c^m_\mathrm{i}$ and $c^m_\mathrm{s}$ is
\begin{equation}
p_m(n) =\frac{1}{u_m^2}\left(\frac{v_m}{u_m}\right)^{2n}= p_m (1-p_m)^n,
\label{eq:SM_proba1mode}
\end{equation}
with $p_m=1-\left(\frac{v_m}{u_m}\right)^2=\frac{1}{u_m^2}=1/G_m$. From this expression, we can compute the average output photon number in mode $m$ of the idler,
\begin{equation}
\langle n\rangle_m=G_m-1,
\label{eq:SM_Average_photon}
\end{equation}
and the corresponding variance,
\begin{equation}
\langle \Delta n^2\rangle_m=G_m(G_m-1).
\label{eq:SM_Variance_photon}
\end{equation}
The probability distribution Eq.~\eqref{eq:SM_proba1mode} can also be rewritten as
\begin{equation}
p_m(n) = \frac{1}{1+\langle n\rangle_m}\left(\frac{\langle n\rangle_m}{1+\langle n\rangle_m}\right)^{n},
\label{eq:SM_proba1modemean}
\end{equation}
where the only parameter is the average photon number in a mode $m$: $\langle n\rangle_m$. We see that the probability of emitting a photon into a mode $m$ knowing that we have the vacuum as an input state follows a thermal distribution with $\langle n\rangle_m=G_m-1$ being the parameter of the law. Importantly, as we expect photons to be emitted in several modes, the resulting photon distribution is expected to be the convolution of several thermal distributions with different $\langle n\rangle_m$. 

\subsection{Measurement of the multimodal vacuum photon distribution}
\begin{figure}[t]
  \centering
  \includegraphics[width=0.48\textwidth]{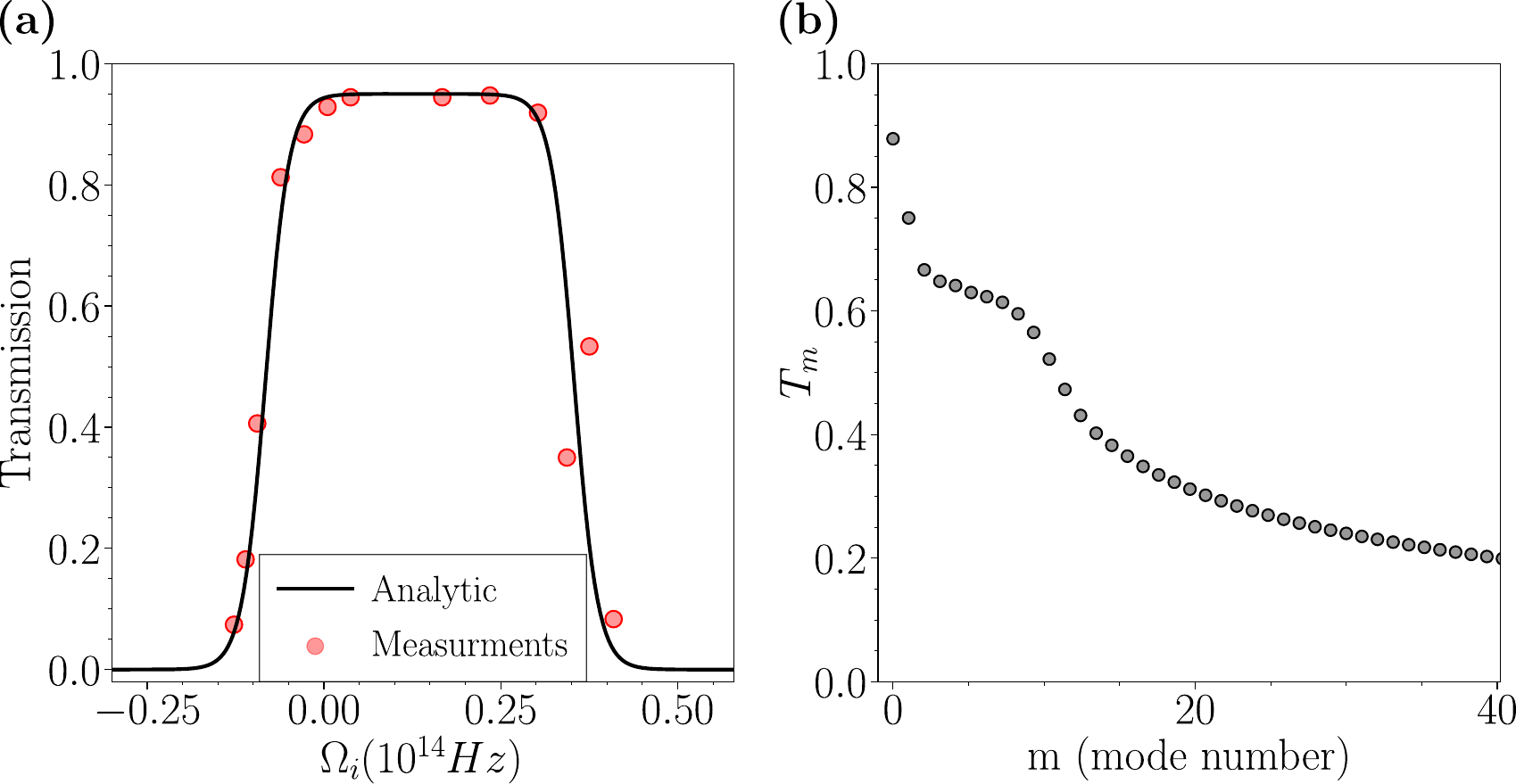}
    \caption{(a) Pass band frequency filter of 13 nm centered at 749 nm. The curve is a fit of the experimental data fitted as a double sigmoid, see text. (b) Calculated transmission of the crystal eigenmode through the filters.}
    \label{fig:SM_FilterandTmode}
\end{figure}

To compare the previous results with experiments, we need, in a first place, to characterize the band-pass filter which is present on the experimental setup and which aims at selecting only a small frequency window around the central idler wavelength. To do so, we model the transmission function of the filter $T(\omega)$ with a high-pass and a low-pass sigmoid functions that we compare with experimental transmission measurements realized with a Ti:Sapph laser whose frequency is varied within a chosen scanning range; see Fig.\ref{fig:SM_FilterandTmode}a). Using this characterization and the explicit form of the mode that we derived numerically, we can deduce the mode transmission by calculating 
 \begin{equation}
     T_m=\int_{-\infty}^{\infty} \phi_m^{*}(\omega)\phi_m(\omega)T(\omega)\text{d}\omega,
     \label{eq:SM_modeTransmiossion}
 \end{equation}
which is shown in Fig.~\ref{fig:SM_FilterandTmode}b). From this transmission coefficient, using the calculation of the previous section, we can deduce the average photon number per mode that would go through the filter and thus will participate in the detection (hits the detector):
\begin{equation}
\langle n\rangle_m^{T}=T_m(G_m-1),
\label{eq:SM_Average_photon_Transmi}
\end{equation}

\begin{figure}[t]
  \centering  \includegraphics[width=0.98\columnwidth]{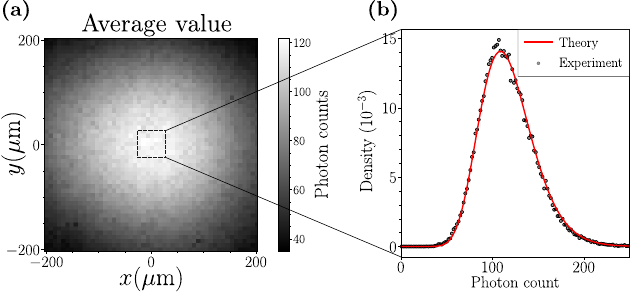}\hfill
    \caption{(a) Average spatial distribution of the measured fluorescence in a pump-only configuration. (b) Experimental photon distribution from pixels in the central region of the image in (a), and theoretical distribution computed with a fitted gain parameter of $g=g_{0}=8.4218\times10^{-13}$ $s.m^{-1}$ (red line).}
     \label{fig:SM_Vacdistribution}
\end{figure}

\begin{figure}[t]
  \centering
  \includegraphics[width=0.24\textwidth]{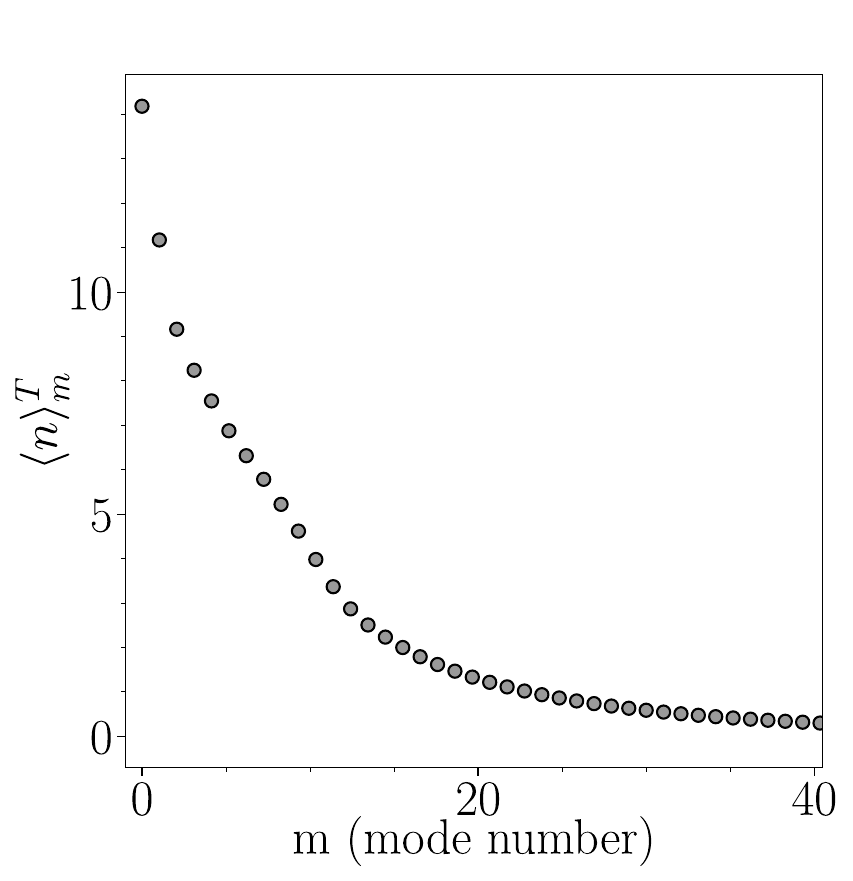}
    \caption{Calculated average photon number per mode emitted by the BBO crystal and passing through the filter.}
     \label{fig:SM_Mean}
\end{figure}
More importantly, we can use the theoretical calculation of the previous section, in association with the experimental characterization of the transmission, to verify if the predictions match experimental measurements of vacuum photon statistics distribution. The measurements are performed by recording 2000 single-shot measurements 
under pump only illumination of the BBO crystal (no input signal field). 
We do the statistical analysis on the pixels located at the center of the measured emission profile to remain in the zone of validity of a plane-wave, 0D approximation (see Fig.~\ref{fig:SM_Vacdistribution}(a)). The histogram of the measured fluorescence after filtering with the use of the spectral filter described above is plotted in Fig.~\ref{fig:SM_Vacdistribution}(b).

Using $g$ as a fitting parameter, we can calculate the joint probability of getting a photon on a detector knowing that different modes emit photons according to their specific law (see Eq.~\eqref{eq:SM_proba1modemean}).
Looking at Fig.~\ref{fig:SM_Mean}, we notice that nearly no photons are emitted in mode $m=40$. In other words,  modes with $m$ larger than 40 are not contributing to the photon statistics and can be disregarded. Thus, to obtain the red curve in the right panel of Fig.~\ref{fig:SM_Vacdistribution}(b) we perform the numerical convolution of the first $40^\text{th}$ thermal laws that follow Eq.~\eqref{eq:SM_proba1modemean} with $\braket{n}^{T}_m$ as a parameter.  We obtain a remarkable agreement in the description of the photon statistics between the experiment and the theory, having only a unique fitting parameter $g=g_{0}=8.4218\times10^{-13}$ s.m$^{-1}$. This agreement validates the temporal multimodal description carried out in this work and shows that it provides a good description to address quantitatively the statistics of photon count emitted by amplified spontaneous emission processes.

\subsection{Calculation of 0D Photon statistics with input coherent state}
\label{SM:0DCoherentstate}

After having explained the behavior of the photon statistics emitted from the vacuum, we aim at describing a simple case of a pure input state which, specifically, in our experimental case, will be a coherent state originating from a Ti:Sapph laser. The first part of this section generalizes the calculation done in Sec.~(\ref{SM:VacuumStat}) to the case of an arbitrary pure input state in the signal (and zero input photon in the idler).

As the temporal modes are independent, the calculation is performed for a single set of modes $m$, thus, the index $m$ is suppressed for clarity. We first compute the distribution of the number of output photons in the idler when starting from $n_0$ input signal photons. Similar results have been obtained in the context of Hawking radiation \cite{Parker1975,Gasperini1990}. We start by computing
\begin{equation}
|0,n_0\rangle_{\rm in} = \frac{(c_{\mathrm{s}}^\dag)^{n_0}}{\sqrt{n_0!}}|0\rangle_{\rm in}
\end{equation}
in terms of the out-states, that is (with $x\coloneqq v/u$, $u=\sqrt{G}$ and $v=\sqrt{G-1}$)
\begin{equation}
|0,n_0\rangle_{\rm in} = \frac{(u b_\mathrm{s}^\dag-vb_\mathrm{i})^{n_0}}{u\sqrt{n_0!}}e^{x b_\mathrm{i}^\dag b_\mathrm{s}^\dag  }|0\rangle_{\rm out}.
\end{equation}
Using 

\begin{equation}
\begin{split}
(u b_\mathrm{s}^\dag-vb_\mathrm{i})e^{x b_\mathrm{i}^\dag b_\mathrm{s}^\dag  }|0\rangle_{\rm out} 
&=(u b_\mathrm{s}^\dag e^{x b_\mathrm{i}^\dag b_\mathrm{s}^\dag  }-v[b_\mathrm{i},e^{x b_\mathrm{i}^\dag b_\mathrm{s}^\dag  }])|0\rangle_{\rm out},\\&=\frac{b^\dag_\mathrm{s}}{u}e^{x b_\mathrm{i}^\dag b_\mathrm{s}^\dag  }|0\rangle_{\rm out},
\end{split}
\end{equation} we have
\begin{equation}
|0,n_0\rangle_{\rm in} = \frac{1}{u^{n_0+1}\sqrt{n_0!}}(b_s^\dag)^{n_0}e^{x b_i^\dag b_s^\dag  }|0\rangle_{\rm out}.
\end{equation}
This implies that the probability of measuring the state $|n,n'\rangle_{\rm out}$ is 
\begin{equation}
|{}_{\rm out}\langle n,n'|0,n_0\rangle_{\rm in}|^2=\delta_{n',n+n_0}\frac{1}{ u^{2(n_0+1)}}\frac{(n+n_0)! }{n_0!n!}x^{2n}.
\label{eq:in_out_fock}
\end{equation}

Considering an initial pure arbitrary state in the signal mode, i.e. 
\begin{equation}
|{\rm ini}\rangle=\sum_{n_0}c_{n_0}|0,n_0\rangle_{\rm in},
\end{equation}
where the $c_{n_0}$ are the superposition coefficients, we can use the results above, and the fact that the states with different $n'$ do not interfere with each other thanks to the Kronecker delta, to directly obtain
\begin{equation}
\begin{split}
P(n)=\frac{1}{u^2 n!}x^{2n}\sum_{n_0}\frac{(n+n_0)! }{n_0!u^{2n_0}}|c_{n_0}|^2.
\end{split}
\end{equation}

If the initial state is a coherent state $|0,\alpha\rangle_{\rm in}$, i.e. $c_{n_0}=e^{-|\alpha|^2/2}\frac{\alpha^{n_0}}{\sqrt{n_0!}}$, then the sum over $n_0$ reads
\begin{equation}
\begin{split}
\sum_{n_0}\frac{(n+n_0)! }{n_0!u^{2n_0}}|c_{n_0}|^2 &= e^{-|\alpha|^2}\sum_{n_0}\frac{(n+n_0)! }{n_0!^2}\left(\frac{|\alpha|^2}{u^2}\right)^{n_0},\\
&=e^{-(1-1/u^2)|\alpha|^2} n! L_n\left(-\frac{|\alpha|^2}{u^2}\right),
\end{split}
\end{equation}
where $L_n(z)$ is the $n$-th Laguerre polynomial, and we used
\begin{equation}
\begin{split}
\sum_{n_0}\frac{(n+n_0)! }{n_0!^2}y^{n_0}&=\partial_y^n \sum_{n_0}\frac{y^{n+n_0}}{n_0!},\\
                        & = \partial_y^n(y^n e^{y}),\\
                        & = e^{y} e^{-y}\partial_y^n(y^n e^{y}),
\end{split}
\end{equation}
as well as Rodrigues' formula for the Laguerre polynomials $L_n(y)=\frac{e^{y}}{n!}\partial_y^n (y^n e^{-y})$.

Therefore, the number distribution for an initial coherent state reads
\begin{equation}
\begin{split}
P^{(\alpha)}(n)=(1-x^2)e^{-x^2|\alpha|^2}x^{2n} L_n\left(-\frac{|\alpha|^2}{u^2}\right).
\label{eq:SM_CoherentProba1mode}
\end{split}
\end{equation}
Using the generating function of the Laguerre polynomials, $\sum_n t^n L_n(y)=\frac{e^{-\frac{ty}{1-t}}}{1-t}$, we obtain the corresponding generating function $Q^{(\alpha)}(z)=\sum_n z^n P^{(\alpha)}(n)$,
\begin{equation}
\begin{split}
Q^{(\alpha)}(z)&=\frac{1-x^2}{1-zx^2}\exp\left(-x^2|\alpha|^2\frac{1-z}{1-zx^2}\right),
\end{split}
\end{equation}
from which we compute the photon number average and variance (e.g. $\langle n\rangle=\partial_zQ^{(\alpha)}|_{z=1}$)
\begin{equation}
\begin{split}
\langle n\rangle&=(|\alpha|^2+1)\frac{x^2}{1-x^2}=(|\alpha|^2+1) v^2,\\
  \langle \Delta n^2\rangle  &=(1+|\alpha|^2(1+x^2))\frac{x^2}{(1-x^2)^2}.
\end{split}
\end{equation}
For a given eigenmode $m$, this can be rewritten in terms of the mode gain $G_m$ and the decomposition of the initial state onto the eigenmode $m$,  $\alpha_m$,
\begin{equation}
\begin{split}
\langle n\rangle_m&=(|\alpha_m|^2+1)(G_m-1),\\
  \langle \Delta n^2\rangle_m  &=G_m(G_m-1)[1+(2-1/G_m)|\alpha_m|^2].
  \label{eq:SM_MeanVarGainDep}
\end{split}
\end{equation}

\begin{figure}[t]
  \centering
  \includegraphics[width=0.48\textwidth]{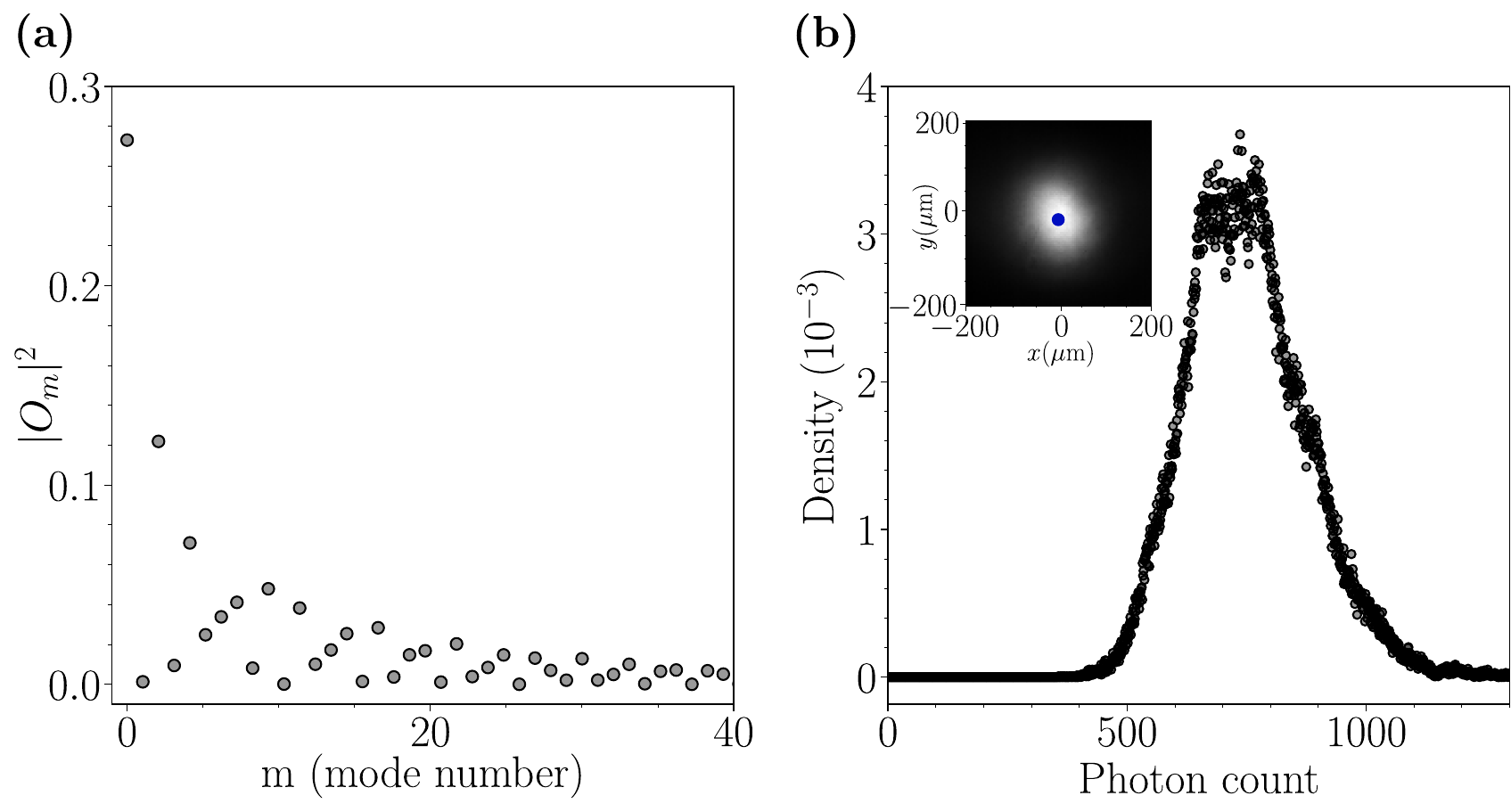}\hfill
	\caption{(a) Square of the overlap integral between the input coherent state and the temporal eigenmodes (b) Normalized histogram of the output photon count measured with 2000 single-shot images.  Inset: average image over 2000 single-shots of the emitted photons spatial profile. The histogram is obtained by selecting the value of the photon on the pixels located at the position of the blue disk.}
     \label{fig:SMHistoSignal}
\end{figure}

At this stage it is important to point that in an ideal single-mode and fluorescence-free response (removing the contribution when $|\alpha|^{2}=0$), the first two moments of the output photon distribution for large Gain ($G\gg1$) are given by

\begin{equation}
\begin{split}
\langle n\rangle_{\text{ideal}}&=GN_0,\\
  \langle \Delta n^2\rangle_{\text{ideal}}  &=2G^{2}N_0,
  \label{eq:SM_MeanVarSinglemod}
\end{split}
\end{equation}
with $N_0=\langle n\rangle_{\text{input}}$ the average photon number in the input coherent state. An input coherent state possesses the generic property that  $\langle n\rangle_{\text{input}}/ \sqrt{\langle \Delta n^2\rangle_{\text{input}}}=\sqrt{N_0}$
that is, its signal-to-fluctuation ratio scales with the square root of the average photon number. In the ideal single-mode response discussed in Eq. \eqref{eq:SM_MeanVarSinglemod}, this ratio becomes $\langle n\rangle_{\text{ideal}}/ \sqrt{\langle \Delta n^2\rangle_{\text{ideal}}}=\sqrt{N_0}/2$. We thus see that the signal-to-fluctuation ratio scales less favorably with the average photon number. This indicates that even in an ideal single-mode scenario, there exists a fundamental limitation arising from vacuum contamination, which increases the relative fluctuations of the output photon distribution by a factor of $\sqrt{2}$. The amplifier does not preserve the coherent photon statistics through amplification processes.  

In order to continue the description using a multimodal approach, it is important to find the $\alpha_m$ coefficients of Eq. \eqref{eq:SM_MeanVarGainDep}. To do so, we need to compute the overlap integral between the initial signal state and the crystal temporal eigenmode $\phi_m$.
For the sake of simplicity, we will assume that the initial state possesses the same spectral limitation as the one imposed by the pump pulse beam: in the experiment, the signal beam is a continuous wave source, and the response of the DFG processes is solely set by the pulsed pump beam. Thus, we will assume an input pulse having a spectrum
\begin{equation}
A_s(\Omega_s)=\exp\left(-4\ln{2} \frac{\Omega_s^{2}}{\Delta_{\omega}^{2}}\right).
\end{equation}
with $\Delta_{\omega}=2.4$ THz as for the pump pulse.
Thus, the overlap is calculated using
\begin{equation}
\text{O}_{m}=\int_{-\infty}^{\infty}  A_{s}(\Omega_s) \phi_{m} (\Omega_s)   \text{d}\Omega_s,
\end{equation}
and shown in Fig.~\ref{fig:SMHistoSignal}(a).

We see that the mode $m=0$ contributes to nearly 30$\%$ of the decomposition. We notice that after $m=20$ the different contributions lie in the percentage level. Finally, the $|\alpha_m|^2$ are just obtained from
\begin{equation}
|\alpha_m|^2=N_{\text{input}} |\text{O}_{m}|^2,
\label{eq:SM_Alpha_m}
\end{equation}
where $N_{\text{input}}$ is the average number of photons in the input state.

\subsection{Measurements of 0D Photon statistics with input coherent state}

To compare the 0D description carried-out previously with the experiments, we need to evaluate the experimental input photon number in the coherent state. We do so by evaluating the number of photons existing over an area corresponding to 1 pixel of the camera ($S=66$ $\mu$m$^{2}$) over a time that include the duration of the pulse pump beam $\Delta_t=2.96$ ps, knowing that the pixel of interest is chosen to be at the central position of the Gaussian intensity beam profile with a waist of $\omega_{r}=140$ $\mu$m and power $P_s$. Thus, the average input photon number at the central pixels of the measured spot is calculated as
\begin{equation}
N_{\text{input}}=\frac{2P_s\Delta_t S}{\hbar \omega_s \omega_r^2\pi}.
\end{equation}

\begin{figure}[t]
  \centering
  \includegraphics[width=0.48\textwidth]{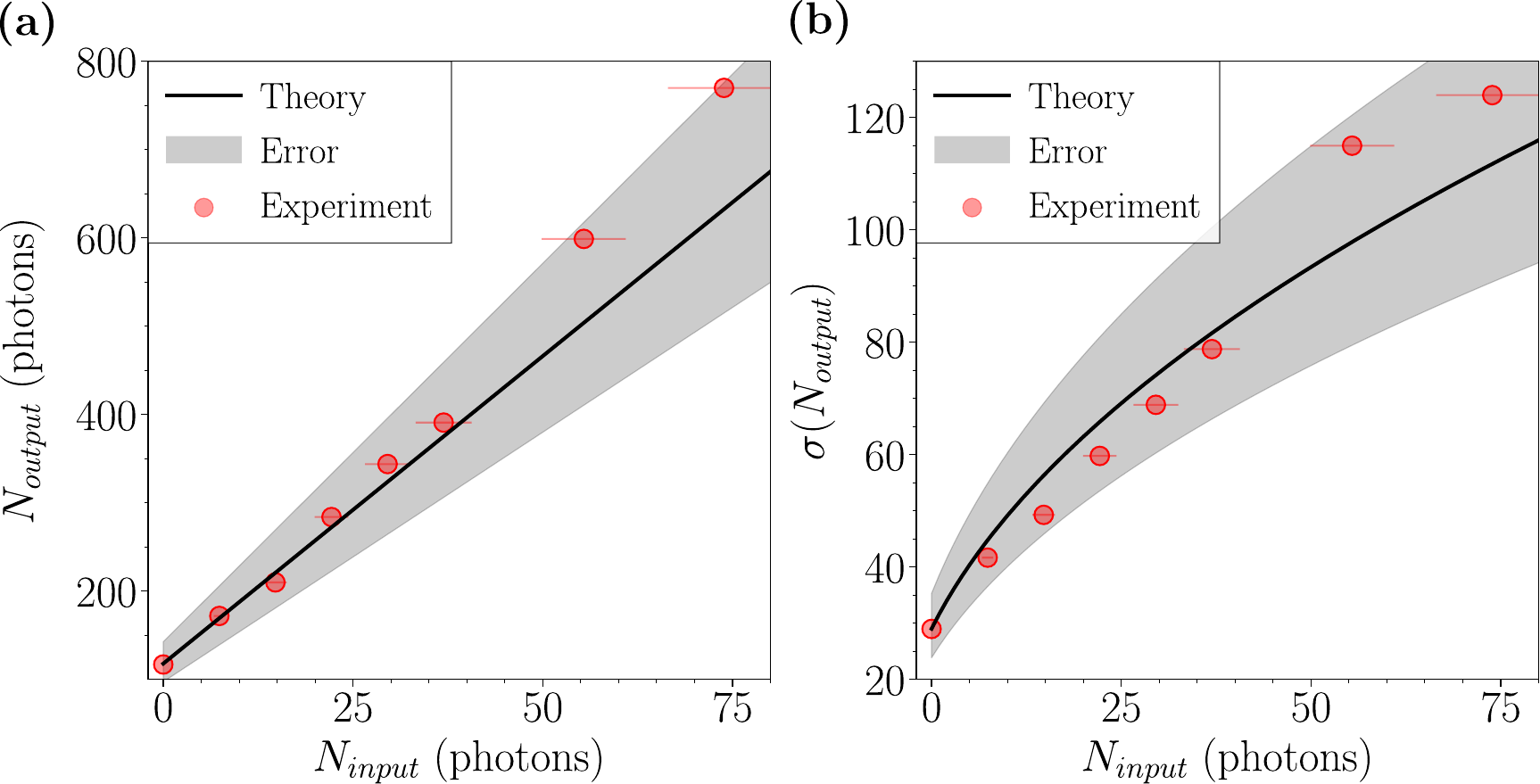}\hfill
  \caption{(a) Average output photon counts as a function of the average input photon number (b) Standard deviation ($\sqrt{\Delta N_{\text{output}}^2}$) of the output photon distribution as a function of the average input photon. The error on the experimental point is a systematic one, estimated to be $\pm 10\%$ due to optical waist and power measurements. The black line is the calculation done with $g_{0}=8.4218\times10^{-13}$ s.m$^{-1}$. The shadow zone around the theory curve shows that prediction if we allow for a variation of $\pm5\%$ of the fitting parameter $g_{0}$.}
    \label{fig:Mean_sigma} 
\end{figure}

We recorded 2000 single-shot images for each $N_{\text{input}}$ that we vary from 0 to 74. An example in the case of $N_{\text{input}}=74$ is given in Fig.~\ref{fig:SMHistoSignal}(b) where the Region Of Interest (ROI) for the analysis is indicated as a blue disk in the inset.
In Fig.~\ref{fig:Mean_sigma} we show the measurement of the first two moments of the photon distributions realized with different power $P_s$ ranging from 0 mW to 2 mW corresponding to a variation of the average input photon number from 0 to 74. The point at $N_{\text{input}}=0$ corresponds to the vacuum case presented in the previous section, where we extracted a fitted value of the $g$ parameter of $g_{0}=8.4218*10^{-13}$ $s.m^{-1}$. Importantly, we keep this gain value estimated from vacuum statistics to compute the prediction with a non-zero signal, using Eq.~\eqref{eq:SM_MeanVarGainDep} to compute the expected photon number emitted as well as the variance contribution of each mode $m$. As the modes are independent, a summation is done over the first $40$ modes to get the total number of photons as well as the variance. Similarly to what has been done for the vacuum statistics, the transmission through the filter is also taken into account.  Finally, we obtain the following formula for the prediction of the average and variance of the output photon number:
\begin{equation}
\begin{split}
N_{\text{output}}&=\sum_{m}T_m(|\alpha_m|^2+1)(G_m-1),\\
\Delta N_{\text{output}}^2&=\sum_{m}T^{2}_{m} G_m(G_m-1)[1+(2-1/G_m)|\alpha_m|^2].
\label{eq:SM_MeanVarsum}
\end{split}
\end{equation}
$|\alpha_m|^2$ is given by Eq.~\eqref{eq:SM_Alpha_m}, $G_m$ by Eq.~\eqref{eq:Gain_lambda_link}, and $T_{m}$ by Eq.~\eqref{eq:SM_Average_photon_Transmi}.
The computed first two moments are plotted with solid lines in Fig.~\ref{fig:Mean_sigma}.
Overall, we find a remarkable agreement between the prediction and experiment, confirming that the system in this configuration behaves effectively as a 0D amplifier whose photon distributions can be fully captured by the multimodal framework introduced above. An essential point is that the gain parameter extracted from vacuum statistics alone is sufficient to reproduce the results obtained with an input coherent state, without any additional fitting. This demonstrates both the predictive power and the internal consistency of the temporal mode approach. Moreover, the analysis highlights the crucial role of the large number of contributing modes --around forty in our case-- in shaping the output photon distribution, underlying how multimodal effects intrinsically limit the preservation of input photon statistics.

\subsection{Calculation of 0D Photon statistics with input thermal state}
\label{SM:ThermalDensityMatrix}
For completeness, we provide here a generalized calculation that describes the case when the input state is a statistical mixture described by a thermal density matrix (keeping the  input idler state as vacuum). Again, as the modes are independent, we perform the calculation on one set of modes and suppress the label $m$ for clarity.

The initial state is 
\begin{equation}
    \rho_{\rm in} = \sum_{n'} \frac{e^{-\beta\hbar\omega n'}}{Z}|0,n'\rangle_{\rm in}{}_{\rm in}\langle 0,n'|,
\end{equation}
with $Z=\sum_{n'}e^{-\beta\hbar\omega n'}$, and $\beta$ the inverse temperature linked to the average photon number. The average number of photons in the input source is $\bar n = (e^{\beta\hbar\omega}-1)^{-1}$, and the probability to measure $n$ photons in the input signal is the usual thermal distribution $\frac{1}{1+\bar n}\left(\frac{\bar n}{1+\bar n}\right)^n$.

The probability of finding $n$ photons in the idler out-state is (using Eq.~\eqref{eq:in_out_fock})
\begin{equation}
\begin{split}
    P^{\text{th}}(n)&= \sum_k {}_{\rm out}\langle n,k|\rho_{\rm in}|n,k\rangle_{\rm out},\\
    &=\sum_{k,n'} \frac{e^{-\beta\hbar\omega n'}}{Z}|{}_{\rm out}\langle n,k|0,n'\rangle_{\rm in}|^2,\\
    &=\sum_{k,n'} \frac{e^{-\beta\hbar\omega n'}}{Z}\delta_{k,n+n'}\frac{1}{u^{2(n'+1)}}\frac{(n+n')!}{n'!n!}x^{2n},\\
    &=\sum_{n'} \frac{e^{-\beta\hbar\omega n'}}{Z}\frac{1-x^2}{n'!}\frac{1}{u^{2n'}}\frac{(n+n')!}{n!}x^{2n},\\
    &=\frac{1-x^2}{Z}\frac{x^{2n}}{n!}\sum_{n'}\frac{(n+n')!}{n'!}\left(\frac{e^{-\beta\hbar\omega}}{u^{2}}\right)^{n'}.
\end{split}
\end{equation}
Using $\sum_{n'} \frac{(n+n')!}{n'!}y^{n'}=\frac{\partial^n}{\partial y^n} \sum_{n'} y^{n+n'}=\frac{\partial^n}{\partial y^n}  \frac{y^n}{1-y}=\frac{n!}{(1-y)^{n+1}}$, we obtain
\begin{equation}
\begin{split}
    P^{\text{th}}(n)&=\frac{1-x^2}{Z}\frac{x^{2n}}{(1-e^{-\beta\hbar\omega}/u^2)^{n+1}},\\
            &=  \frac{1-e^{-\beta \hbar\omega}}{u^2-e^{-\beta \hbar\omega}}\left(\frac{v^2}{u^2-e^{-\beta \hbar\omega}}\right)^n,\\
            &= p_{\text{th}}(1-p_{\text{th}})^n,
\end{split}
\end{equation}
defining $p_{\text{th}}$. We remark right away that this distribution is also thermal, with mean photon number  $\langle n\rangle=\frac{1-p_{\text{th}}}{p_{\text{th}}}=\frac{v^2}{1-e^{-\beta \hbar\omega}}=(G-1)(\bar n+1)$ and 
\begin{equation}
P^{\text{th}}(n)=\frac{1}{1+\langle n\rangle}\left(\frac{\langle n\rangle}{1+\langle n\rangle}\right)^n.
\label{eq:thermaldistrib}
\end{equation}

Putting back the label $m$, we find that the output mean and variance of the idler photon number in mode $m$ reads
\begin{align}
\langle n\rangle_m &= (G_m-1)(\bar n_m+1), \label{eq:nmean}\\
\langle \Delta n^2 \rangle_m 
   &= (G_m-1)(\bar n_m+1)  \bigl[(G_m-1)(\bar n_m+1)+1\bigr] \nonumber\\
   &= \langle n\rangle_m(\langle n\rangle_m+1).
   \label{eq:nvariance}
\end{align}
The latter result is that of a thermal distribution. Importantly, we see that if the system is set in single mode configuration, a thermal input state will stay thermal (with a higher temperature) after the amplification. This is in stark contrast to the coherent case when, in a single-mode restriction, we have seen that the coherent photon statistic is not preserved after amplification.

\begin{figure}[t]
  \includegraphics[width=0.48\textwidth]{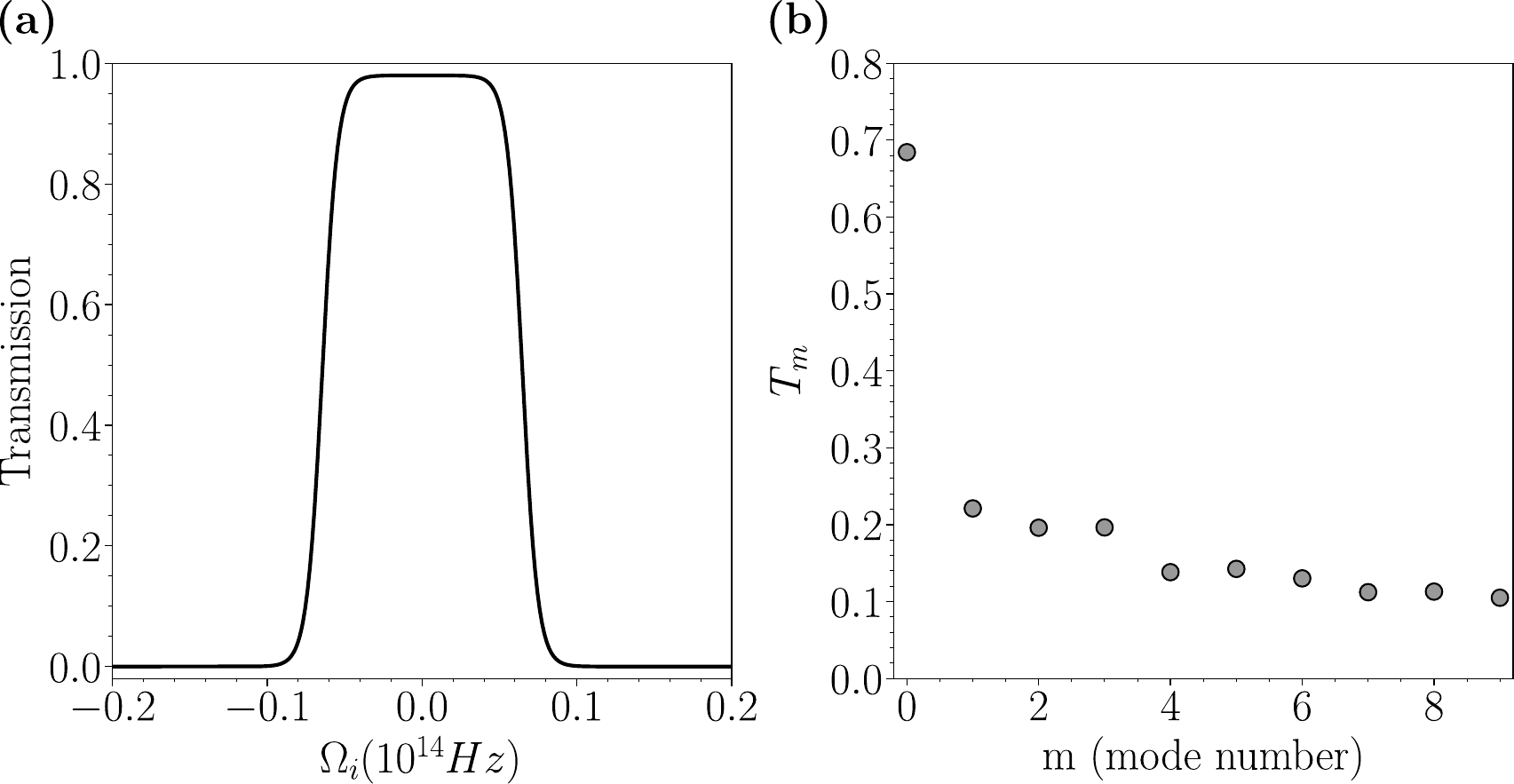}\hfill
    \caption{(a) Filter transmission in the configuration optimized for $\lambda_s=$840nm. (b) Computed mode transmission $T_m$.}
    \label{fig:SM_FIlter2D_Vac} 
\end{figure}

\section{Experimental settings for 2D measurements}
\label{SM:2Dconfiguration}

At the beginning of the manuscript, we have presented two-dimensional measurements allowing us to distinguish qualitatively between two different photon number statistics. We have seen that the measurements do not show quantitatively the ideal behavior associated with thermal and coherent photon statistics. In this section, we will use the theoretical results displayed in Sec.~\ref{sec:multimodal} to derive predictions that capture the experimental behavior and, thus, explain the observed alteration of the photon statistics through the nonlinear amplifier.

We start by explicitly providing all the information about the parameters used to perform the measurements on the photon distributions of a coherent and thermal source, which were different compared to the 0D characterization developed in Sec.~\ref{sec:multimodal}. Indeed, to improve the imaging capability of the experimental setup, we changed the BBO crystal with a $l_c=2$ mm length, allowing for a larger phase-matching bandwidth for the transverse $k$ component of the image of interest, and we doubled the spatial size of the pulse pump beam to increase the constant gain region in the transverse plane.
In addition, as we want to obtain statistical information on the photons, we narrow down the frequency filter window to maximize the contribution originating mainly from one mode. The filter is configured as presented in Fig.~\ref{fig:SM_FIlter2D_Vac}(a), which leads to the mode transmission presented in Fig.~\ref{fig:SM_FIlter2D_Vac}(b). 
The phase-matching is optimized for a signal of $840$ nm and an idler of $763.6$ nm (this corresponds to an angle of 29.14 degrees). 

\begin{figure}[t]
  \includegraphics[width=0.48\textwidth]{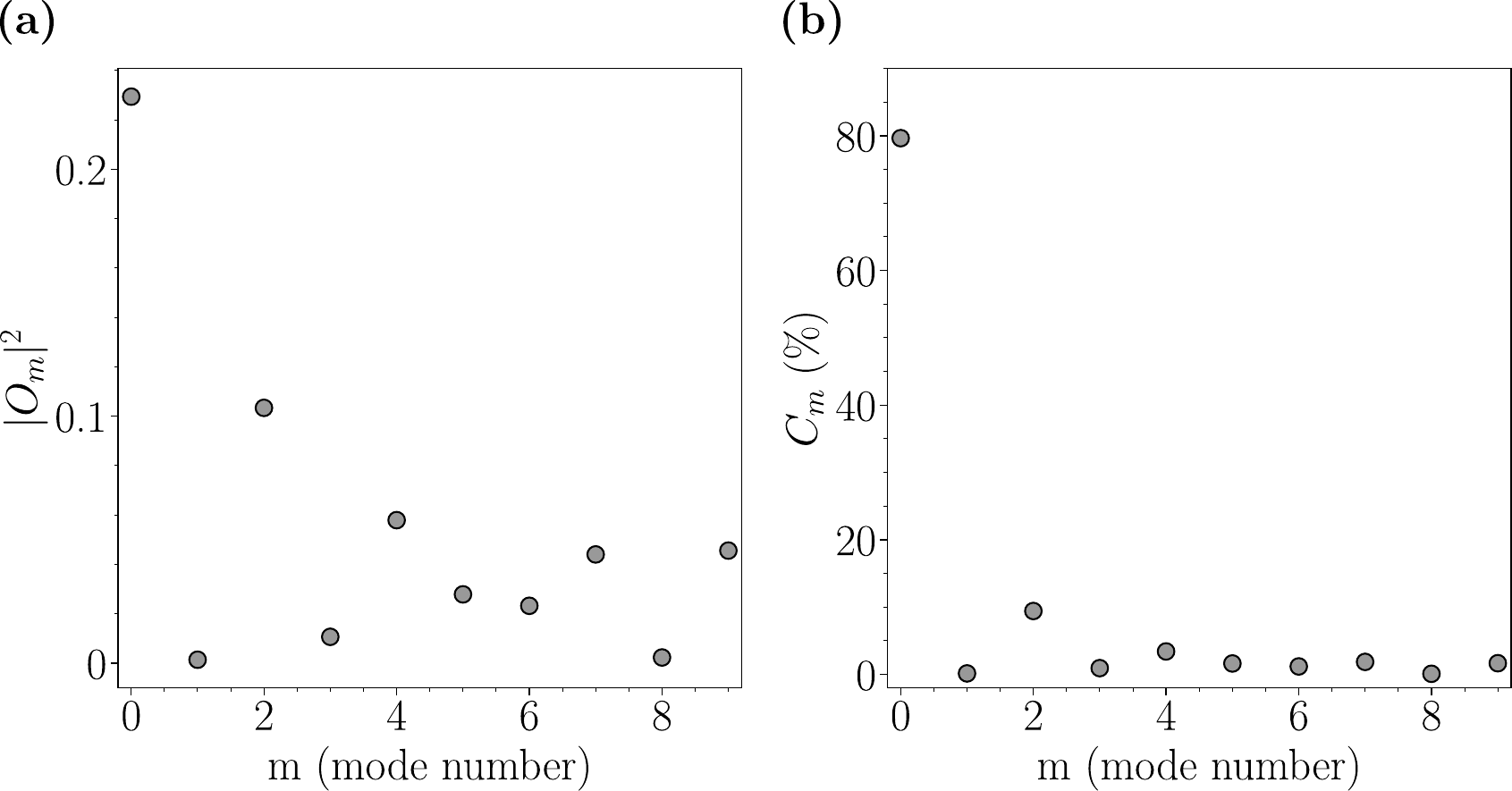}\hfill
    \caption{(a) Calculated mode overlap between the initial state and the temporal mode of the system when optimized for a signal wavelength of $840$ nm and a filter in the configuration of Fig.~\ref{fig:SM_FIlter2D_Vac}.
    (b) Computed relative contribution of each mode to the statistical information at the output, Eq.~\ref{eq:SM_Contribution}.}
    \label{fig:SM_TransmOverlap2D} 
\end{figure}

We do not present here the full details of the calculation of the singular values and the temporal modes, as they are conceptually the same as presented in Sec~\ref{sec:multimodal}, with a crystal length of $l_c=2$ mm and dispersion parameters presented in table \ref{fig:SM_tab:dispersion2D}. We keep the same FWHM for the pump spectrum as in Sec~\ref{sec:multimodal}: $\Delta_{\omega}=2.4$ THz.
From the new set of eigenmodes, we calculate the mode overlap with the input state accounting for the pump pulse duration, see Fig.~\ref{fig:SM_TransmOverlap2D}(a).

\begin{table}[t]
\footnotesize
\begin{tabular}{l|c}
\hline\hline
 & $\lambda_{\mathrm{p}}=400\operatorname{nm}$, $\lambda_{\mathrm{s}}=840\operatorname{nm}$, $\lambda_{\mathrm{i}}=763,6\operatorname{nm}$ \\
\hline
$k'_{\mathrm{p}}\,(\times10^{-9}\operatorname{s\,m^{-1}})$ & $5.81385$ \\
$k'_{\mathrm{s}}\,(\times10^{-9}\operatorname{s\,m^{-1}})$ & $5.61079$ \\
$k'_{\mathrm{i}}\,(\times10^{-9}\operatorname{s\,m^{-1}})$ & $5.62757$ \\
\hline
$k''_{\mathrm{p}}\,(\times10^{-25}\operatorname{s^{2}\,m^{-1}})$ & $1.96116$ \\
$k''_{\mathrm{s}}\,(\times10^{-26}\operatorname{s^{2}\,m^{-1}})$ & $6.8984$ \\
$k''_{\mathrm{i}}\,(\times10^{-26}\operatorname{s^{2}\,m^{-1}})$ & $8.0428$ \\
\hline
\end{tabular}
\caption{Dispersion parameters for critical phase-matching of BBO at room temperature optimized with an angle of 29.14 degree.}
\label{fig:SM_tab:dispersion2D}
\end{table}

To analyze the statistical distribution of the output photons, it is useful to define a quantity that isolates the contribution of each temporal mode to the amplified signal of interest. We therefore introduce the average photon number per mode, corrected for fluorescence background, which reads
\begin{equation}
N^{m}_{\text{int}}=T_m N_{\text{input}}|O_{m}|^{2} (G_{m}-1).
\label{eq:SM_MeanInterest}
\end{equation}

From this definition, one can directly evaluate the relative weight of each mode in carrying the statistical information of the input signal. This is quantified by
\begin{equation}
C_{m}=\frac{N^{m}_{\text{int}}}{\sum_{m}N^{m}_{\text{int}}},
\label{eq:SM_Contribution}
\end{equation}
which informs us how much each temporal mode contributes to the transmission of the input photon statistics through the amplifier. A large value of $C_{m}$ indicates that most of the statistical information is carried by a single dominant mode, whereas smaller values distributed over many modes indicate that the information is spread across a multimodal response. 
Figure~\ref{fig:SM_TransmOverlap2D}(b) displays the computed values of $C_m$. We see that the first mode contributes to 80$\%$ and the third one to 10$\%$. The remaining contribution is split into the other modes. We designed the experiment to produce this response, highly peaked on the first mode, by selecting a small filtering window. 
The reason is that if we aim at preserving in the output field the statistical properties of the input field as much as possible, we would need to work only with 1 temporal mode. 
Otherwise, if we have a large collection of modes that contribute significantly to the output, the measured statistics will be a collection of the emitted photons in all the modes, which will be described by the joint probability of all the independent mode distributions.
The latter, in the limit of a large mode number, will finally result in a Gaussian distribution due to the central limit theorem and, thus, information on the input field statistics will be washed out in the amplification process.

We can now compute the expected photon distributions at the output of the BBO crystal using as input states the coherent or thermal sources used in the experiments and displayed in Figs.~\ref{fig:AmeanVar} and~\ref{Fig:DoubleHisto}. It will allow us to compare the multimodal approach used in this article with the experimental observations.
To do so, we compute the first 10 modal contributions to the idler output states using the parameters presented in this section and let the value of $g$ as a free parameter. 
We use the explicit formula for the photon distribution given by Eqs.  \eqref{eq:thermaldistrib} and \eqref{eq:SM_CoherentProba1mode} for the thermal and coherent statistics, respectively. 
In the case of the thermal distribution, the average number of photons per mode that serve as the parameter of the law is given by $\braket{n}_m=(G_m-1)N^{\text{th}}_{\text{input}}|O_m|^{2}$ with $N^{\text{th}}_{\text{input}}=35$. In the case of the coherent distribution, each mode has $|\alpha_m|^{2}=N^{\text{ch}}_{\text{input}}|O_m|^{2}$ and $x^{2}_m=(1-1/G_m)$ with $N^{\text{ch}}_{\text{input}}=51.5$. 
The gains, $G_m$, calculated from Eq.~\eqref{eq:Gain_lambda_link} are given by the singular values of the SVD, the crystal length $l_c$ of 2mm, and a fitted value of $g=g_{2D}=5.963\times10^{-13}$ $s.m^{-1}$. 
The smaller fitted value of $g$ compared to the experiments discussed in Sec.~\ref{sec:multimodal} is consistent with the fact of having an experimentally larger pump beam size in the case of the measurements of Figs.~\ref{fig:AmeanVar} and~\ref{Fig:DoubleHisto}, while keeping the power constant. For each mode $m$, knowing $T_m$, we calculate the expected photon distribution after the filter. To be complete, we also consider the false count distribution coming from our detector. In our case, the ORCA fusion camera has readout noise distributed in a Gaussian manner with a standard deviation of $\sigma=3.05$ photon count. 

The calculated probability distributions for thermal and coherent input signals are plotted in Fig. \ref{Fig:DoubleHisto}. The predictions obtained from the multimodal framework reproduce the experimental photon distributions with excellent accuracy. This agreement confirms that the deviations observed from ideal coherent and thermal statistics originate naturally from the combined effects of vacuum contamination (fluorescence) and the multimode response of the amplifier, which offers a clear explanation of the experimental system's response.

 \section{CONCLUSION AND OUTLOOK}   

In this work, we have demonstrated a method to measure spatially resolved photon statistics using difference-frequency generation (DFG) in a BBO crystal. By introducing a temporal multimode decomposition of the system response, we have shown that the amplification process can be described as a set of independent parametric channels, each corresponding to a temporal eigenmode. This framework provides an accurate and physically intuitive modeling of both fluorescence generation and signal amplification, and explains how multimode effects shape the measured photon statistics.

The central result of this study is twofold. First, we establish the capability of resolving photon statistics across two spatial dimensions and demonstrate that our setup can qualitatively discriminate between coherent and thermal input states. Second, we show that the observed deviations from ideal distributions can be fully explained by vacuum contamination and the multimodal response of the amplifier. This dual outcome both validates the experimental platform and highlights the importance of temporal modes in determining the fidelity of photon statistics measurements.

Beyond this proof of principle, our approach establishes a robust optical platform for bidimensional photon statistics measurements. Such a capability is particularly relevant for hybrid light–matter systems, where photon correlations encode key signatures of many-body physics. In exciton–polariton microcavities, for instance, single-shot access to photon statistics in two dimensions could provide crucial insights into the onset of superfluidity \cite{Carusotto2013,Amo2009,Amo2009b,Amo2011,Sanvitto2010,Fontaine2022,Caputo2017} , the role of fluctuations, and the emergence of turbulent behavior\cite{Claude2020,Amelio2020} driven by vortex interactions.  

Looking ahead, several promising directions emerge from this work. First, the integration of engineered nonlinear materials, such as periodically poled crystals, could offer increased flexibility in phase-matching and allow tailoring of the temporal mode structure. Considering Type-II phase-matching could offer twofold benefits: the possibility of separating signal and idler based on polarization, and the opportunity to engineer the process to operate in a temporally single-mode regime. Finally, other nonlinear processes could also be considered for one-shot two-dimensional photon statistics measurements, such as sum-frequency generation or four-wave mixing. The use of SFG could allow direct access to the input statistics, without the need to infer them as we have done here, provided that the $100\%$ efficiency limit is reached. While in single-pass configurations, such as those in \cite{Eckstein2011, Christ2013, Reddy2013},  the efficiency cannot exceed $80\%$, temporal-mode interferometry schemes, such as those proposed by \cite{Reddy2014}, could allow approaching near-unity efficiency. 

\section*{ACKNOWLEDGMENTS}   
This work was supported by the European Research Council grant EmergenTopo (865151), the QuantERA project MOLAR funded by the Agence National de la Recherche (ANR-24-QUA2-0006), the projects ANR-24-CE30-6695 FUSIoN, ANR-24-CE47-4949 UniQ-RingS and ANR-23-PETQ-0001 Dyn1D at the title of France 2030, and by the European Union, the French government through the Programme Investissement d’Avenir (I-SITE ULNE /ANR16-IDEX-0004 ULNE) managed by the Agence Nationale de la Recherche, the Labex CEMPI (ANR-11LABX-0007) and the region Hauts-de-France (CPER WAVETECH). This project has received funding from the European Union’s Horizon 2020 research and innovation programme under the Marie Sklodowska-Curie grant agreement No 101108433. 
\section*{DATA AVAILABILITY}  
The data that support the findings of this article are openly available \cite{DataAv}.
\newpage
\section*{APPENDIX: SINGLE-SHOT MEASUREMENTS}   

\begin{figure}[h]
	\centering
	\includegraphics[width=1\linewidth]{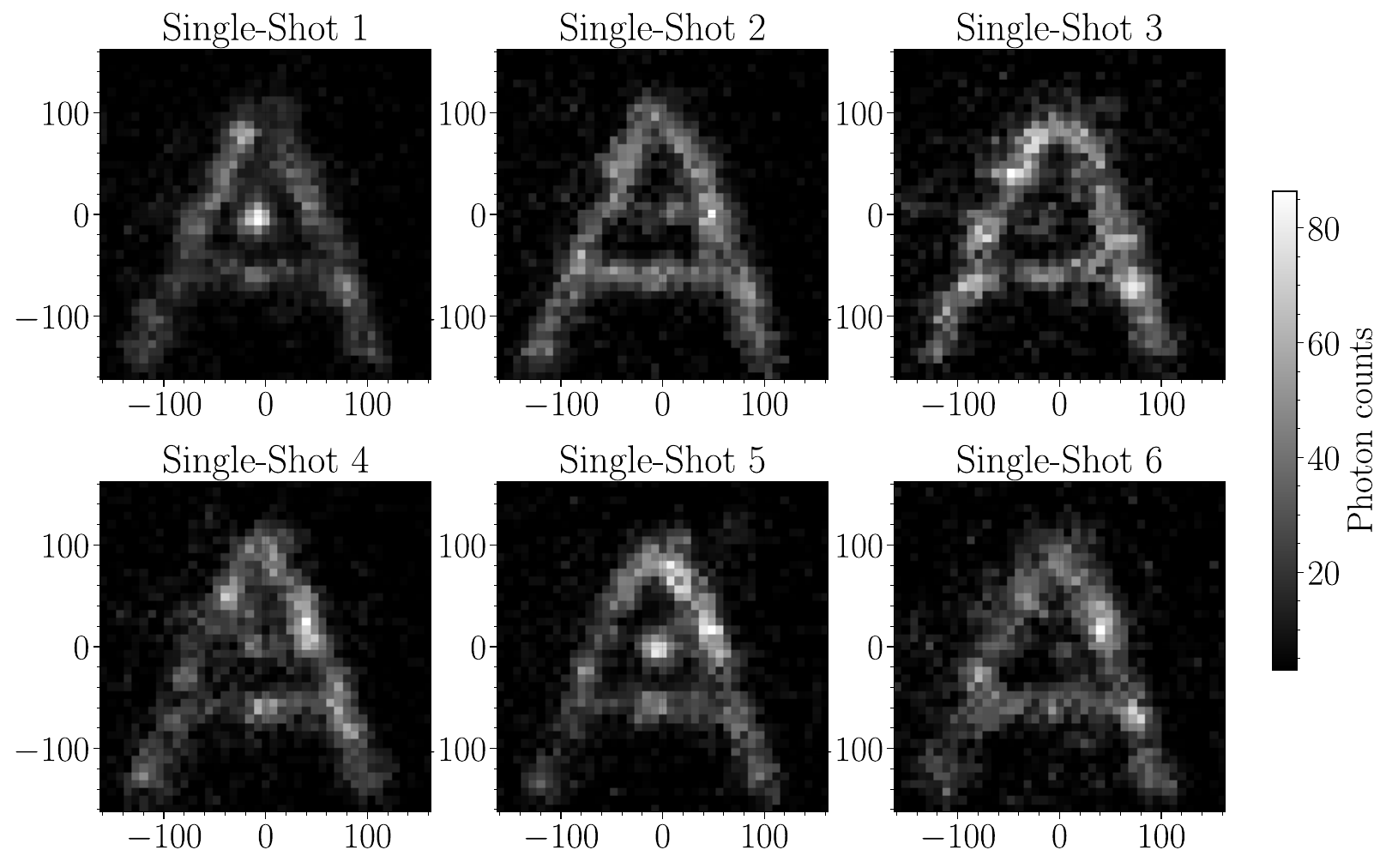}
	\caption{Single-shots of a 2D spatial distributions. The A shape is made from a SLM and a Ti:Sapph laser, while the central spot originated from a filtered Exalos light source. The axis are in microns.}
\label{fig:Sm_Singleshot} 
\end{figure}

For completeness, in Fig.\ref{fig:Sm_Singleshot}) we present a few single shots from the pool of 10000 recorded images.

\section*{APPENDIX: SPECTRAL AND TEMPORAL PROPERTIES OF THE PUMP FIELD}
\label{SM:pumpspectrum}

The intense $400$ nm pulse used for DFG is generated by frequency doubling an 800 nm pulse emitted from a Legend Elite Coherent laser system. The temporal width (FWHM) of the 800 nm pulse was measured with an autocorrelator (APE PulseCheck) and found to be $\Delta_t^{\text{autoco}} = 4.2\ \mathrm{ps}$. Assuming a Gaussian intensity profile, we apply a deconvolution factor of 0.707, yielding an intensity FWHM of $\Delta_t^{I,800\text{nm}} = 0.707 \times \Delta_t^{\text{autoco}} = 2.96\ \mathrm{ps}$. In the experiment, this pulse is frequency-doubled to produce the 400 nm beam. Ideally, the intensity FWHM of the resulting 400 nm pulse is expected to be reduced by a factor $\sqrt{2}$, giving $\Delta_t^{I,400\text{nm}} = \Delta_t^{I,800\text{nm}}/\sqrt{2} = 2.02\ \mathrm{ps}$. Correspondingly, the FWHM of the 400 nm field amplitude envelope is expected to be $\Delta_t = \sqrt{2} \times \Delta_t^{I,400\text{nm}} = 2.96\ \mathrm{ps}$.
\\

Nevertheless, we measured the spectrum of the 400 nm pulse with an Optical Spectrum Analyzer (OSA) and found a FWHM of $\Delta_{\omega}=2.4$ $\mathrm{THz}$. For information, the Fourier limit due to the pulse duration for a Gaussian pulse is $\Delta^{F}_{\omega}=0.44\times 2\pi/\Delta_t=0.93$ $\mathrm{THz}$. The measured spectral width is thus approximately three times larger than this limit. This discrepancy likely arises from the frequency-doubling process used to generate the 400 nm pulse, which can introduce spectral phase distortions. As we do not have precise information about the resulting spectral phase, and since its impact is limited to minor modifications of the temporal mode shape and fitted gain values, we choose to neglect it. For simplicity, we will therefore assume throughout the paper that the pump pulse spectrum is given by: 

\begin{equation}
A^{u}_{\text{p}}(\Omega_p=\omega-\omega_p)=A^{0}_{\text{p}}A_{\text{p}}(\Omega_p),
\end{equation}

with $\Omega_p$ the sideband of the pump. By energy conservation ($\Omega_p=\Omega_s+\Omega_i$), we can rewrite the pump spectrum like: $A_{\text{p}}(\Omega_i+\Omega_s)=\exp{-[4\ln{2} \frac{(\Omega_i+\Omega_s)^{2}}{\Delta_{\omega}^{2}}]}$ with $\Omega_i=\omega_1-\omega_i$, $\Omega_s=\omega_2-\omega_s$, $A^{0}_{\text{p}}$ the unitless peak spectral amplitude. $\omega_1$ and $\omega_2$ are two independent frequencies necessary to understand the nonlinear processes
\section*{APPENDIX: SLM SETUP}   
\label{SM:SLMdetails}
\begin{figure}[h!]
        \centering
        \includegraphics[width=1\linewidth]{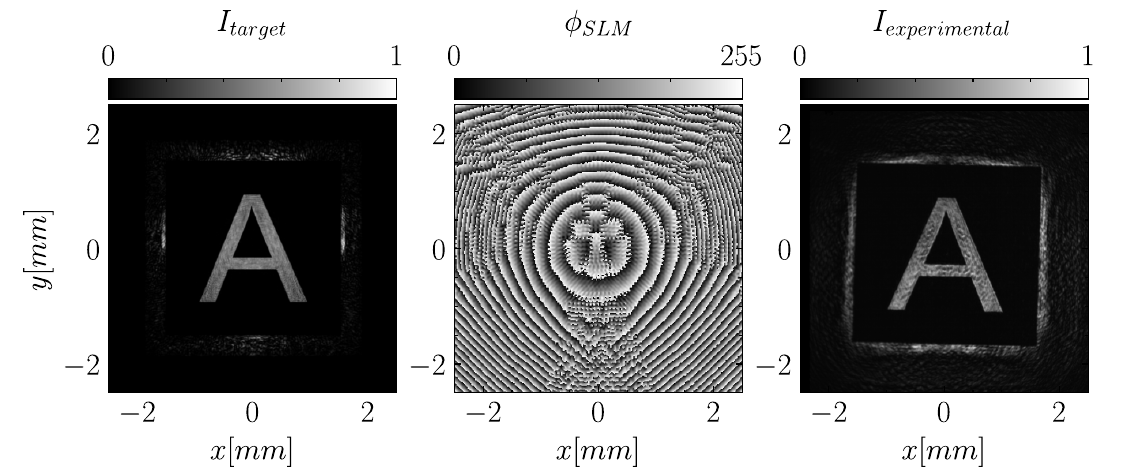}
        \caption{\textbf{(a)} Target intensity pattern generated by the MRAF algorithm with mixing parameter m=0.4 after n= 100 iterations.\textbf{(b)} Corresponding phase mask generated and addressed on the 
        as shown in Fig. (\ref{Fig:setup}). \textbf{(c)} Experimentally measured intensity distribution at the Fourier plane of an $f=150$ mm lens.}
        \label{fig:SM_SLM_A}
\end{figure}

This part provides a details about the creation of a complex input image using an SLM.
The A input single image used in the second part of the manuscript to demonstrate faithful imaging was created by tailoring the coherent input beam from the Ti:Sapph laser with a phase-only Spatial Light Modulator (\href{https://holoeye.com/products/spatial-light-modulators/pluto-2-1-lcos-phase-only-refl}{HOLOEYE Pluto 2.1 LCOS}).
Doing so we created a complex shape possessing a rich transverse $k_{x}$ and $k_{y}$ distribution. To realize this input state, we developed a homemade MRAF algorithm\cite{MRAF_pasienski08} allowing us to obtain a desired image with $20\%$  overall error.
The explicit Python code developed for creating the phase mask can be found in the corresponding GitHub repository \cite{ouahrouche_SLM}. The target intensity, the phase mask applied to the SLM, as well as the recorded image measured with a regular camera at the Fourier plane of the SLM, are represented in Fig. \ref{fig:SM_SLM_A}.
\section*{APPENDIX: CRYSTAL GAIN AND RELATION WITH EXPERIMENTAL PARAMETERS}
\label{SM:gain}

We have seen above that the mode gain is defined by 
\begin{equation}
G_{m}=\cosh^{2}{(g \lambda_m l_c)}.
\end{equation}
with $l_c$ the crystal length over the propagation axis and $g$ is linked to the medium and fields through \cite{LoudonBook}
\begin{equation}
g=|A^{0}_{\text{p}}|\frac{\chi^{(2)}}{2}\sqrt{\frac{\omega_s\omega_i}{c^{2} n(\omega_s)n(\omega_i)}},
\end{equation}
where $n(\omega_s)$ and $n(\omega_i)$ are the crystal refractive index evaluated at signal and idler frequencies respectively, and $\chi^{(2)}=3.91\times10^{-12}$ m.V$^{-1}$ the nonlinear coefficient associated to the BBO crystal. $A^{0}_{\text{p}}$ is the spectral amplitude of the pump. In the experiment, $A^{0}_{\text{p}}$ depends on the waist and the optical power of the pump beam, which can vary shot-to-shot. That is why in the main plot of the paper, we allowed for a variation of $\pm5\%$ of $g$ to obtain the shadow area that illustrates the sensitivity to typical experimental variations of $g$.



\bibliography{sample}

\end{document}